\RequirePackage{fix-cm}
\PassOptionsToPackage{hyperfootnotes=false}{hyperref}
\documentclass[fleqn,usenatbib]{mnras}
\usepackage{newtxtext,newtxmath}
\usepackage{multirow}
\usepackage{multicol}
\usepackage{makecell}

\usepackage{caption}
\usepackage{subcaption}  
\usepackage{placeins}
\usepackage{float}
\usepackage{natbib}

\usepackage[T1]{fontenc}
\usepackage{microtype}
\DeclareRobustCommand{\VAN}[3]{#2}
\let\VANthebibliography\thebibliography
\def\thebibliography{\DeclareRobustCommand{\VAN}[3]{##3}\VANthebibliography}

\usepackage{graphicx}	
\usepackage{amsmath}	

\title[Statistical analysis of Neutron Star Cooling
]{Statistical Analysis of Neutron Star Cooling: \\ Implications for the Nuclear Equation of State and Nucleon Pairing}

\author[A. Ávila et al.]{
Afonso Ávila,$^{1,2,3}$\thanks{afonso.avila@student.uc.pt}
Ricardo Z. Ferreira$^{1,4}$\thanks{rzferreira@fisica.uminho.pt}
and Violetta Sagun$^{2}$\thanks{v.sagun@soton.ac.uk}
\\
\\
$^{1}$CFisUC, Department of Physics, University of Coimbra, Rua Larga P-3004-516, Coimbra, Portugal\\
$^{2}$Mathematical Sciences and STAG Research Centre, University of Southampton, Southampton SO17 1BJ, United Kingdom\\
$^{3}$Observatoire Astronomique de Strasbourg, Université de Strasbourg,
CNRS, 11 rue de l’Université, 67000 Strasbourg, France\\
$^{4}$Centro de F\'isica das Universidades do Minho e do Porto (CF-UM-UP), Universidade do Minho, P-4710-057 Braga, Portugal \\
}

\pubyear{2026}

\begin{document}
\label{firstpage}
\pagerange{\pageref{firstpage}--\pageref{lastpage}}
\maketitle

\begin{abstract}
We perform a statistical analysis of the thermal evolution of isolated neutron stars by confronting theoretical cooling curves with luminosity–age measurements for 24 sources. For each object, we explore five purely nucleonic equations of state—APR4, FSU2R, DD2, IST, and NL3$\omega\rho$—combined with representative models of neutron superfluidity and proton superconductivity, while varying the gravitational mass and envelope composition. The analysis spans approximately $10^5$ cooling curves. Within this framework, we find that only the NL3$\omega\rho$ equation of state and, to a lesser extent, IST - both including neutron triplet pairing—provide an overall satisfactory description of the data. The remaining equations of state are in significant tension with the observations. These results demonstrate that NS cooling provides complementary constraints on dense matter, alongside nuclear structure, global NS observables, and gravitational-wave observations, while probing the interplay between the EoS and nucleon pairing.

\end{abstract}


\section{Introduction}
\label{sec:Introduction}

The extreme densities reached in the interior of neutron stars (NSs) provide a natural laboratory for studying strongly interacting matter beyond the reach of terrestrial experiments. NS properties are determined by the equation of state (EoS) of dense matter, which relates the pressure, density, and energy density and thus allows astrophysical observations to constrain the underlying microphysics. Different observables probe different aspects of the EoS. Precision measurements of massive pulsars already rule out very soft EoSs~\citep{Antoniadis:2013pzd,Romani:2021xmb}. Constraints on the NS mass–radius relation are provided by X-ray observations with NICER~\citep{Miller:2019cac,Vinciguerra:2023qxq,Dittmann:2024mbo,Salmi:2024aum,Salmi:2024bss,Choudhury:2024xbk,Mauviard:2025dmd} and by multi-messenger detection of the binary NS merger GW170817, which probes the tidal deformability~\citep{LIGOScientific:2018cki,LIGOScientific:2020aai,Dietrich:2020efo,Kunert:2026hmx}. 
In addition, the thermal evolution of NSs provides a complementary probe of dense matter, as cooling predictions for a given EoS can be directly confronted with the X-ray observations of isolated NSs
~\citep{Blaschke:2004vq,Popov:2004ey,Blaschke:2006gd,Wei:2018yaj,Wei:2020xdk,Fortin:2021umb,Das:2024ams,Marino:2024gpm}, and an opportunity to test physics beyond the Standard Model~\citep{Sedrakian:2015krq,Sedrakian:2018kdm,Hamaguchi:2018oqw,Buschmann:2021juv,AngelesPerez-Garcia:2022qzs,Avila:2023rzj,Giangrandi:2024qdb}.

The standard cooling scenario involves thermal photon emission from the stellar surface and neutrino emission from the NS interior. After the early thermal-relaxation time, during which the crust and core become thermally coupled, NSs cool predominantly through neutrino emission from the bulk of the star and, at later times, through photon emission from the surface~\citep{Page:2004fy,Yakovlev:2004iq,Potekhin:2015qsa}. The efficiency of neutrino emission processes is sensitively dependent on the particle composition. In particular, once the kinematic threshold set by the Fermi-momentum triangle condition for the $\beta$-decay process and its inverse, known as the direct Urca (DU) process, is reached, the stellar temperature drops rapidly~\citep{Lattimer:1991ib}. If this condition is not met, the less efficient modified Urca (MU) process, which involves an additional spectator nucleon and has no threshold, along with nucleon–nucleon bremsstrahlung, results in slower cooling~\citep{Page:2004fy}.

Another key factor affecting the thermal evolution of NSs is the formation of Cooper pairs. According to the Bardeen–Cooper–Schrieffer (BCS) theory~\citep{Cooper:1956zz,Bardeen:1957mv,Bohr:1958zz}, an attractive interaction between degenerate fermions leads to the formation of Cooper pairs. Neutrons form singlet-state ($^1S_0$) superfluidity in the inner crust, but this pairing is suppressed in the core, where the effective neutron–neutron interaction in the singlet channel becomes repulsive. Instead, core neutrons are expected to pair in the triplet state ($^3P_2$), while protons can undergo singlet-state pairing~\citep{Potekhin:2015qsa}. Pairing opens an energy gap $\Delta$ in the single-particle excitation spectrum around the Fermi surface, with a width of $2\Delta$. These paired phases affect NS thermal evolution through two competing mechanisms. On the one hand, it suppresses standard neutrino emission processes and reduces the heat capacity. On the other hand, it opens an additional neutrino emission channel associated with Cooper pair breaking and formation (PBF), which can enhance cooling when the temperature drops below the critical temperature~\citep{Baym:1969fiw,Chen:1986xzn,Baldo:1992kzz,Yakovlev:1999sk,Yakovlev:2004iq,Sedrakian:2018ydt}.
While nucleon singlet-state pairing is well-studied, the formation of the neutron $^3P_2$ channel in the NS core remains uncertain~\citep{Lombardo:2000ec,Dean:2002zx}; therefore, cooling observations provide a sensitive probe to the nucleon pairing and gap models. 

For a given EoS, envelope composition, and stellar mass, the predicted time evolution of the surface temperature or luminosity can therefore be directly compared with observations, allowing one to assess which nuclear properties are compatible with the data. In this work, we focus on the set of 24 isolated NSs (see Appendix~\ref{app:obs_app}) of different ages~\citep{PotekhinCoolData} and perform a global statistical analysis by comparing a large set of theoretical cooling curves, generated with the public code \textsc{NSCool}~\citep{Neutronstarcoolingcode}, to the available luminosity age data. We consider five purely nucleonic EoSs, APR4, DD2, FSU2R, IST, and NL3$\omega\rho$, and a representative set of neutron and proton pairing models. For each EoS and pairing gap combination, we scan over the NS gravitational mass and envelope composition, and quantify the agreement with the data via a chi-square analysis. 

Among the observed sources, the compact object at the center of the Cassiopeia~A (Cas~A) supernova remnant is of particular interest because of its young age and the rapid decrease in its surface temperature. With an estimated age of approximately 356 years, it is one of the youngest known isolated NSs.
~\citep{Ho:2009mm,Heinke:2010cr,Shternin:2010qi,Elshamouty:2013nfa,Wijngaarden:2019tht,Shternin:2021fpt}. A recent combined re-analysis of the X-ray spectra suggests a surface temperature decrease of 1.4-2.5\% over two decades of observations and a mass estimate of $M= 1.55\pm0.25\ \mathrm{M_\odot}$~\citep{Shternin:2022rti} offering a sensitive probe for rapid cooling mechanisms and nucleon pairing in the stellar core~\citep{Page:2010aw,Blaschke:2011gc,Blaschke:2013vma,Ho:2014pta,Taranto:2015ubs}. It is a subject of debate due to its unusually rapid temperature decrease, which has motivated extensive discussion of possible rapid cooling mechanisms, including a strong neutron $^3P_2$ pairing in the core~\citep{Page:2010aw,Ho:2014pta}, rapid cooling driven by activation of the DU process in middle-mass stars~\citep{Taranto:2015ubs}, modified neutrino emission via a one-pion exchange~\citep{Blaschke:2011gc,Blaschke:2013vma}, etc. 

The paper is organized as follows. Section~\ref{sec:eoss} describes the EoSs used and NS properties. Section~\ref{sec:nscooling} summarizes the neutrino and photon emission processes from NSs and the nuclear pairing combinations adopted. Section~\ref{sec:stats} describes the statistical analysis, and the results are given in Section~\ref{sec:Results}. We present our conclusions in Section~\ref{sec:Conclusion}. The observational data are given in Appendix~\ref{app:obs_app}.

Throughout the article, we use natural units with $\hbar = c = 1$, except for the stellar luminosity, which is expressed in $\mathrm{erg/s}$.

\section{NS Properties}
\label{sec:eoss}
\subsection{Hadronic EoSs}
To account for the different baryonic matter uncertainties, we use five different models, the APR4, IST, FSU2R, DD2, and NL3$\omega\rho$ EoSs, for $\beta$-equilibrated matter, with different compositions, stiffness, and saturation density properties. In Table~\ref{tab:astro_params} we list their main physical properties and briefly summarize each in the following.
\begin{center}
\captionsetup{type=table}
\begin{minipage}{\columnwidth}
\centering
\begin{tabular}{lcccc}
\hline
EoS &
\makecell{$\mathrm{M_{max}}$\\ \mbox{[$\mathrm{M_\odot}$}] } &
\makecell{$\mathrm{R_{1.4}}$\\ \mbox{[km]}} &
\makecell{$\mathrm{M_{DU}}$\\ \mbox{[$\mathrm{M_\odot}$]}} &
\makecell{$\mathrm{n_{DU}}$\\ \mbox{[fm$^{-3}$]}} \\
\hline
APR4             & 2.17 & 11.33 & 1.98 & 0.78 \\
DD2              & 2.42 & 13.20 & N/A  & N/A  \\
FSU2R            & 2.05 & 12.80 & 1.92 & 0.61 \\
IST              & 2.08 & 11.40 & 1.91 & 0.87 \\
NL3$\omega\rho$  & 2.75 & 13.82 & 2.55 & 0.50 \\
\hline
\end{tabular}
\caption{Parameters of the considered models. The columns correspond to the maximum gravitational mass, radius of the 1.4 $\mathrm{M_\odot}$ star, gravitational mass, and baryonic density at the onset of the nucleonic DU process for each EoS considered.}
\label{tab:astro_params}
\end{minipage}
\end{center}
The {\bf APR4} EoS~\citep{Akmal:1998cf} is based on the Argonne v18 two-nucleon interaction~\citep{Wiringa:1994wb,Machleidt:1995km} supplemented by relativistic boost corrections and the three-body forces~\citep{Wiringa:1988tp}. It is known for its comparatively soft behaviour at supranuclear densities and for including microscopic many-body correlations~\citep{Akmal:1997ft}. The model achieves realistic saturation properties and supports a maximum mass of about 2 $\mathrm{M_\odot}$ when three-body interactions are included.

The {\bf DD2} EoS~\citep{Typel:2009sy} is a nucleonic relativistic mean–field (RMF) EoS with density-dependent meson-nucleon couplings calibrated to nuclear structure data and constrained by microscopic Dirac-Brueckner calculations at low densities. The DD2 EoS provides a relatively stiff symmetry energy and yields large radii. An important feature of the DD2 EoS is its consistent treatment of medium effects through density-dependent couplings, enabling it to reproduce both finite-nucleus properties and homogeneous matter up to several times saturation density.

The {\bf FSU2R} EoS~\citep{Tolos:2016hhl} is an updated version of the RMF FSU2 model~\citep{Chen:2014sca}, predicting smaller NS radii while satisfying the observational constraint on the maximum NS mass. The main modifications concern the density dependence of the nuclear symmetry energy, with a reduction in its slope from $L=113$ MeV in FSU2 to $L=44.3$ MeV in FSU2R. This softens the pressure of neutron-rich matter around nuclear saturation density, leading to more compact NS configurations. At the same time, the EoS retains sufficient stiffness at higher densities to support a maximum mass of approximately $2.05~M_\odot$. The FSU2R parameters are constrained by empirical nuclear data, ensuring a consistent description of finite nuclei and bulk nuclear matter.

The {\bf IST} EoS~\citep{Sagun:2018cpi} is based on the induced surface tension (IST) framework, which accounts for the short-range hard-core repulsion between particles. The IST term arises naturally in dense matter due to short-range repulsion and provides a smooth transition between the excluded volume description at low densities and the eigenvolume regime at high densities, where the IST contribution vanishes~\citep{Sagun:2013moa}. In the dilute-gas limit, the IST EoS accurately reproduces the first four virial coefficients of a gas of hard spheres. The hard-core radius of nucleons is constrained by fits to particle ratios measured in heavy-ion collision experiments over a wide range of center-of-mass collision energies~\citep{Sagun:2017eye}, while the eigen surface tension parameters are determined to reproduce the nuclear liquid–gas phase transition and its critical endpoint~\citep{Sagun:2016nlv,Kucherenko:2021oxg}. The long-range attraction of the mean-field type and symmetry energy parameters are constrained by NS observations and the tidal deformability inferred from the GW170817 binary NS merger~\citep{Sagun:2018cpi}. In this work, we employ Set~B of the IST EoS developed in~\citep{Sagun:2020qvc}.

The {\bf NL3$\omega\rho$} EoS~\citep{Horowitz:2001ya} is an extension of the original NL3 RMF EoS~\citep{Lalazissis:1996rd} in which the density dependence of the symmetry energy is modified through an isoscalar-isovector $\omega$–$\rho$ mixing term. This additional nonlinear coupling alters the effective $\rho$–meson mass allowing for variations in the symmetry energy slope while preserving the well-established isoscalar properties of the NL3 model and its successful description of finite nuclei. The resulting family of parametrizations is consistent with current constraints and yields a relatively stiff high-density behaviour, supporting NS gravitational masses above 2 $\mathrm{M_\odot}$~\citep{Carriere:2002bx,Pais:2016xiu}. In this work, we used the NL3$\omega\rho$ formulation presented in~\citep{Pais:2016xiu}, which satisfies current nuclear matter and astrophysical constraints.

It is worth noting that the particle composition of the adopted EoSs differs. While the IST EoS includes neutrons, protons, and electrons, the APR4, FSU2R, DD2, and NL3$\omega\rho$ EoSs also include muons. The presence of muons modifies the DU threshold by increasing the minimum proton fraction required to satisfy momentum conservation. As a result, the DU process is activated at higher gravitational masses~\citep{Page:2004fy}.

For the realistic description of the outer layers, all the EoSs are supplemented by the Haensel-Zdunik (HZ) EoS for the outer crust and the Negele-Vautherin (NV) EoS for the inner crust~\citep{Negele:1971vb,1990A&A...227..431H}.
\begin{table}
\centering
\begin{tabular}{lccccc}
\hline
Channel & Model & \makecell{$T_{c,\mathrm{max}}$\\ \mbox{[$10^{9}$\,K]}} & \makecell{$k_{F,0}$\\ \mbox{[fm$^{-1}$]}} & \makecell{$\delta k_F$\\ \mbox{[fm$^{-1}$]}} & \makecell{$r$\\ \mbox{[ \, \, ]}} \\
\hline
           & N1 (SFB) & 5.108 & 0.810 & 0.564 & 1.086 \\
$n\,^1S_0$ & N2 & 5.416 & 0.731 & 0.424 & 0.663 \\
           & N3 (CCDK) & 5.724 & 0.652 & 0.284 & 0.240 \\
\hline
           & T1 (a) & 1.000 & 1.800 & 0.500 & 0.000 \\
$n\,^3P_2$ & T2 & 5.500 & 2.150 & 0.600 & 0.000 \\
           & T3 (c) & 10.000 & 2.500 & 0.700 & 0.000 \\
\hline          
           & P1 (AO) & 2.350 & 0.490 & 0.310 & 0.000 \\
$p\,^1S_0$ & P2 & 4.475 & 0.575 & 0.415 & 0.345 \\
           & P3 (CCDK) & 6.600 & 0.660 & 0.460 & 0.690 \\
\hline
\end{tabular}
\caption{Free parameters describing the Gaussian profile of the pairing gap for the neutron singlet (n $^1S_0$), neutron triplet (n $^3P_2$), and proton singlet (p $^1S_0$) channels. See text for details.}
\label{tab:ourpairings}
\end{table}

\subsection{Nuclear superfluidity and superconductivity}
\label{sec:super}
\begin{figure*}
\centering
\setkeys{Gin}{width=2.00\columnwidth}
\includegraphics{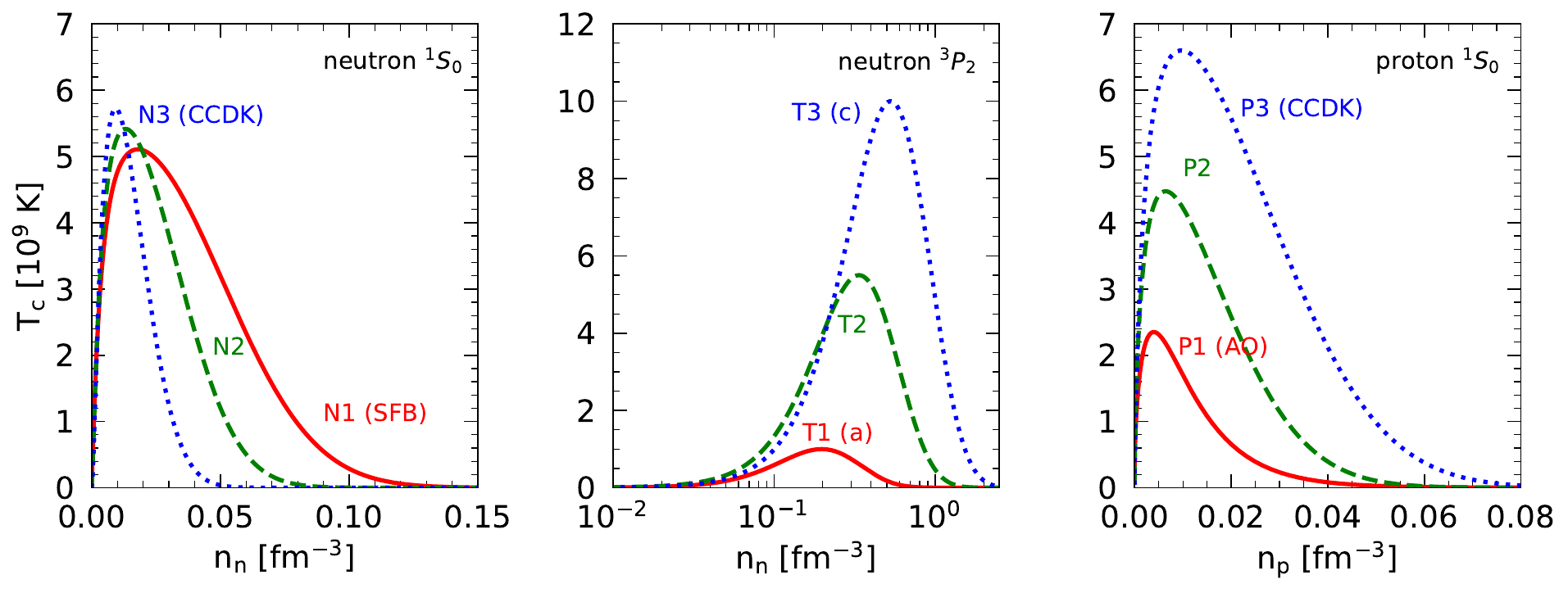}
\caption{Critical temperatures of singlet neutron (left panel), triplet neutron (middle panel), and singlet proton (right panel) superfluidities in NS matter, as functions of the neutron number density, $\mathrm{n_n}$, and the proton number density, $\mathrm{n_p}$, for different superfluidity and superconductivity models for the parameters used in Table~\ref{tab:ourpairings}.}
\label{fig:pairings}
\end{figure*}

Nucleon pairing in the crust and core affects the thermal evolution of NSs by modifying the heat capacity and neutrino emissivity. While superfluidity and superconductivity suppress standard neutrino emission processes, Cooper pair breaking and formation (PBF) provide an additional neutrino emission channel that can enhance cooling~\citep{Lattimer:1991ib,Page:2005fq,Potekhin:2015qsa}.

Given the significant uncertainties associated with nucleon pairing gaps~\citep{Lombardo:2000ec,Dean:2002zx}, particularly for the neutron triplet pairing channel, it is often convenient to describe it by simple analytic functions. In this work, we adopt the parametrization proposed by~\cite{Yanagi:2019zne}, based on a modified Gaussian function
\begin{equation}
    T_c(k_F) = T_{c,\mathrm{max}}\exp\!\left[-\left(\frac{k_F - k_{F,0}}{\delta k_F}\right)^{\!2}-r\left(\frac{k_F - k_{F,0}}{\delta k_F}\right)^{\!4}\right],
\end{equation}
where $T_{c,\mathrm{max}}$ denotes the maximum critical temperature predicted by the model, $k_{F,0}$ is the Fermi momentum at which this maximum is obtained, $\delta k_F$ represents the width of the Gaussian profile in momentum space, and $r$ is an additional dimensionless free parameter introduced to adjust the shape of the pairing gap. All these quantities are treated as parameters of the model.

In total, nine models were considered, with three Gaussian parametrisations for each of the three pairing channels: neutron singlet ($n ^1S_0$), proton singlet ($p ^1S_0$), and neutron triplet ($n ^3P_2$). For the $n ^1S_0$ channel, parametrisations ‘N1’ and ‘N3’ correspond to the SFB~\citep{Schwenk:2002fq} and CCDK~\citep{Chen:1993bam} models, respectively. In the $p ^1S_0$ channel, ‘P1’ and ‘P3’, correspond to the AO~\citep{Amundsen:1984hua} and CCDK~\citep{Chen:1993bam} pairing models, while for the $n ^3P_2$ channel, ‘T1’ and ‘T3’, they represent the ‘a’ and ‘c’ models, respectively~\citep{Kaminker:2001ag,Kaminker:2001eu}. The ‘N2’, ‘P2’ and ‘T2’ models were constructed as intermediate cases, with parameters taken as the arithmetic means of the corresponding values in the two limiting models. For the triplet supplemented pairing, we only take permutations in the singlet neutron and proton channels of the most representative models, namely ‘N1’ and ‘N3’ for the $n ^1S_0$ and ‘P1’ and ‘P3’ for the $p ^1S_0$, and combine each of them with the three $n ^3P_2$ ‘T1’, ‘T2’ and ‘T3’. A cross-check of the Gaussian fits against the pairing models already implemented in \textsc{NSCool} confirmed full consistency with our fitted results. Altogether, we consider each pairing combination as an ordered set `$i$-$j$-$k$', with $i$, $j$, and $k$ denoting the identifiers of the pairing models. Table~\ref{tab:ourpairings} lists the different pairing and parameter values considered in this work. 

\subsection{TOV equations}
\label{sec:tov}
\begin{figure}
\centering
\setkeys{Gin}{width=0.95\columnwidth}
\includegraphics{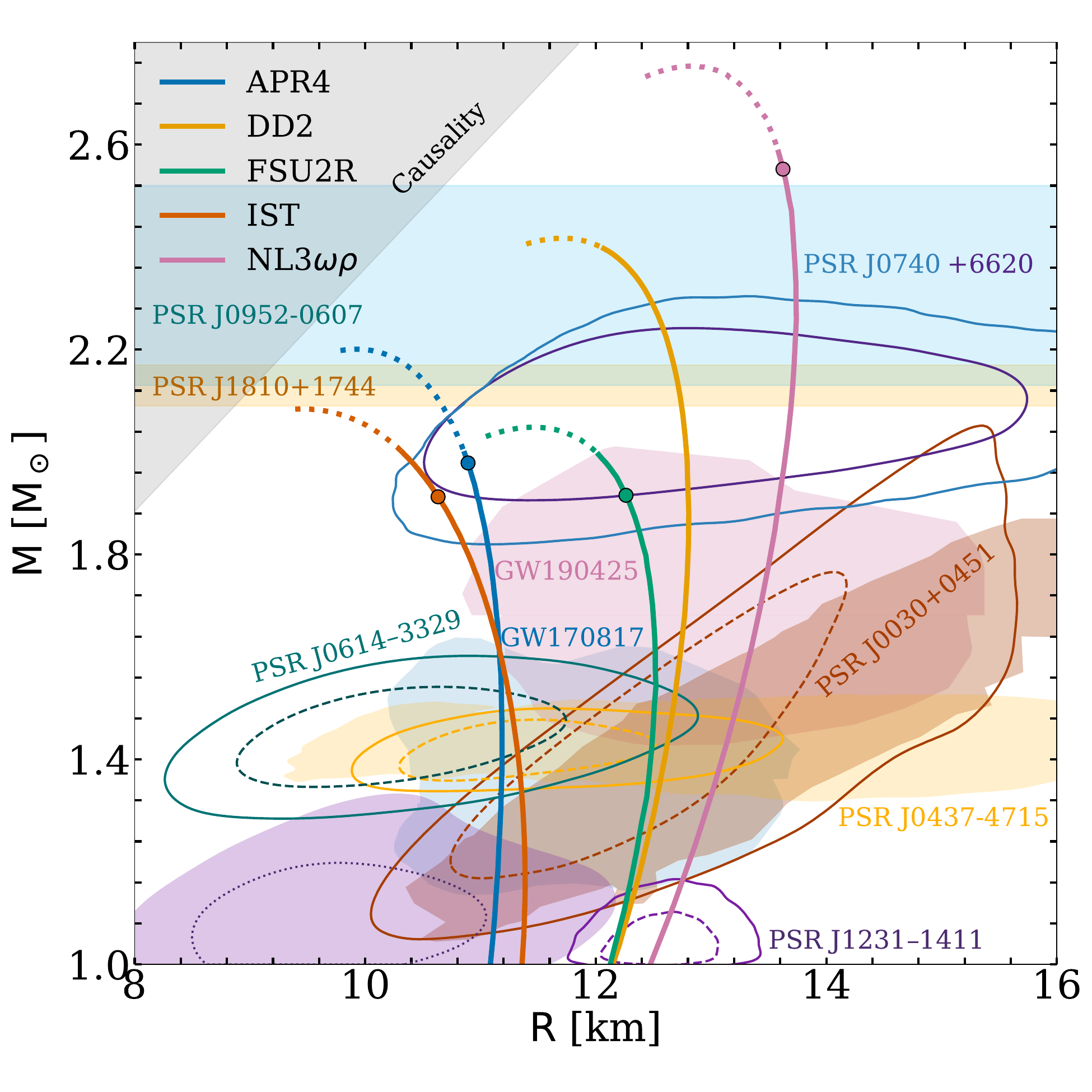}
\caption{Gravitational mass as a function of the star radius for the considered EoSs. The filled circles denote the DU onset mass for each EoS. The dotted segments indicate regions not covered by the simulations, while the solid segments correspond to the mass values included in the cooling dataset. The yellow and cyan bands represent the $1\sigma$ mass constraints on the heaviest pulsars PSR J1810+1744~\citep{Romani:2021xmb}, and PSR J0952-0607~\citep{Romani:2022jhd}, respectively. The updated NICER measurement for PSR J0030+0451 is shown by the solid (95\% CL) and dashed (68\% CL) brown contours~\citep{Kini:2026rjx}, in addition to the filled region (95\% CL)~\citep{Miller:2019cac}. PSR J0437-4715 is depicted with dashed (68\% CL) and solid (95\% CL) yellow contours~\citep{Choudhury:2024xbk}, together with the yellow filled region~\citep{Miller:2025qfq}. The PSR J0740+6620 constraints are shown by the solid purple~\citep{Dittmann:2024mbo} and blue~\citep{Salmi:2024aum} contours. The results from the LIGO-Virgo Collaboration for the GW170817~\citep{LIGOScientific:2018cki} and GW190425~\citep{LIGOScientific:2020aai} binary NSs mergers are shown in teal and magenta, respectively. The solid and dashed dark green contours show the NICER measurements of PSR J0614-3329~\citep{Mauviard:2025dmd}. PSR J1231-1411 is depicted with the purple filled region and the dotted contour inside it, corresponding to the 95\% and 68\% CL~\citep{Qi:2025mpn}, and with the solid (95\% CL) and dashed (68\% CL) purple contours~\citep{Salmi:2024bss}. The gray-shaded region in the top-left corner is ruled out by the causality constraint.}
\label{fig:mr}
\end{figure}

The astrophysical properties of NSs are obtained by solving the Tolman-Oppenheimer-Volkoff (TOV) equations, which describe hydrostatic equilibrium in general relativity for a spherically symmetric and non-rotating configuration~\citep{Oppenheimer:1939ne,Tolman:1939jz}. For a given EoS, these equations determine the radial profiles of pressure and enclosed mass as functions of the energy density
\begin{equation}\label{TOV}
    \frac{dp}{dr}=-\frac{G(\varepsilon +p)(M+4\pi r^3p)}{r^2\left(1-{2GM}/{r}\right)},
\end{equation}
\vspace{-0.5cm}
\begin{equation}
    \frac{dM}{dr}=4\pi r^2\varepsilon.
\end{equation}
Here, $\varepsilon$ and $p$ denote the energy density and pressure, respectively, while $M(r)$ is the gravitational mass enclosed within radius $r$. The gravitational mass and radius are obtained by integrating the TOV equations outward from the centre, where the energy density is fixed as an initial condition, until the pressure vanishes at the surface. The gravitational mass as a function of radius for all the EoSs considered is shown in Fig.~\ref{fig:mr}.
\section{NS Cooling}
\label{sec:nscooling}
\begin{figure*}
    \centering
    \begin{subfigure}[t]{0.32\textwidth}
        \centering
        \includegraphics[width=\linewidth]{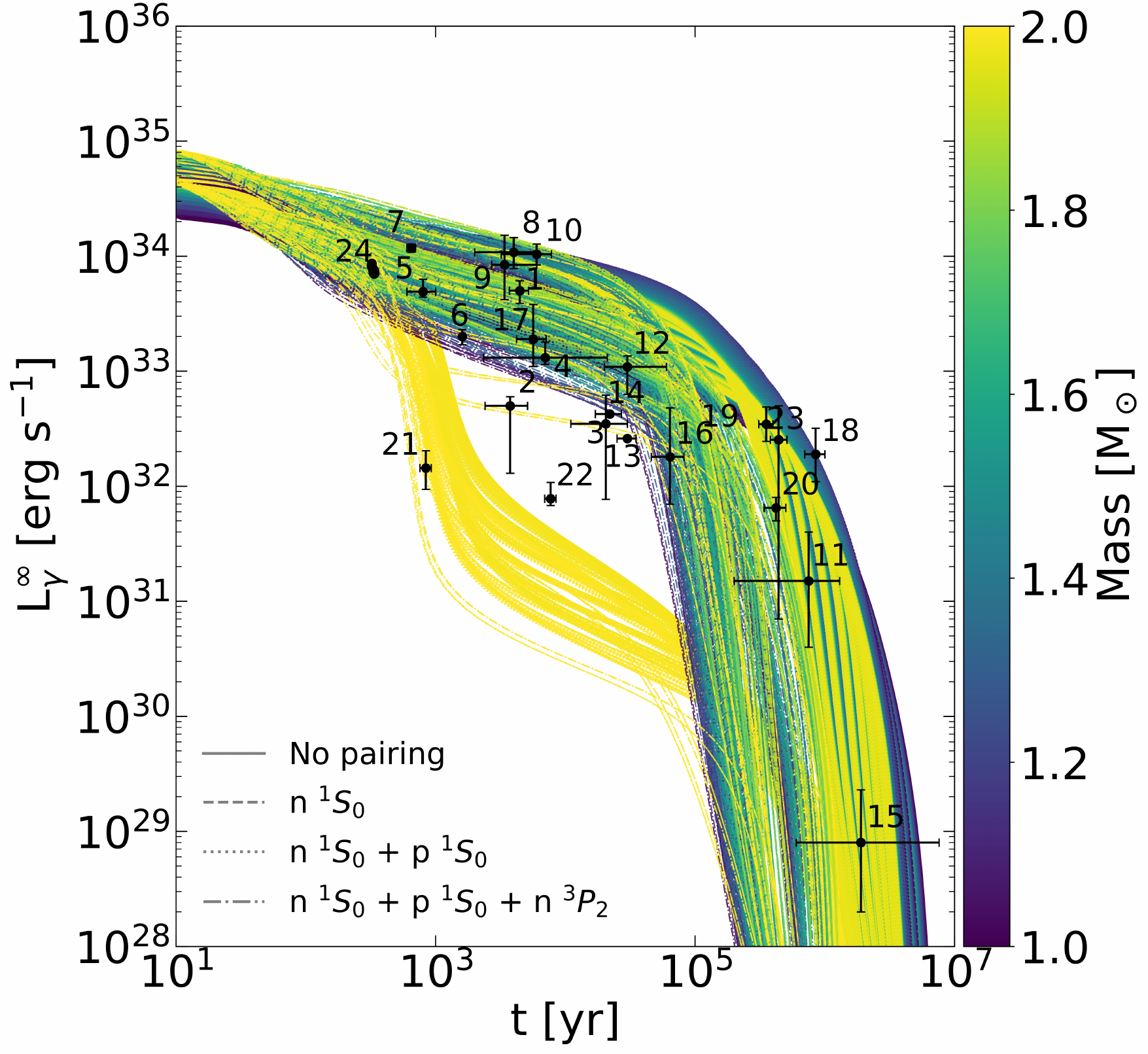}
        \caption{APR4}
        \label{fig:eos_panel_apr4}
    \end{subfigure}\hfill
    \begin{subfigure}[t]{0.32\textwidth}
        \centering
        \includegraphics[width=\linewidth]{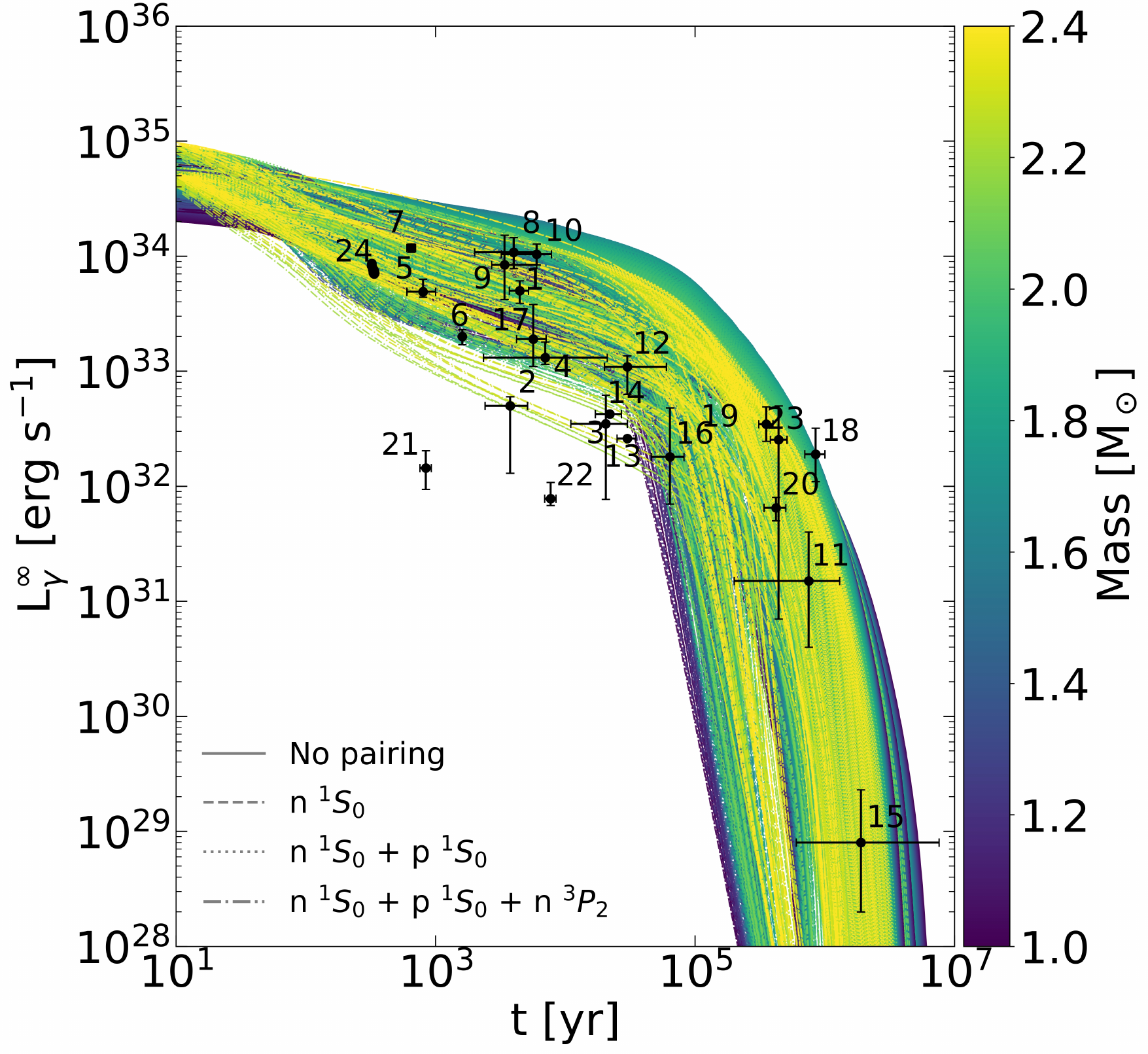}
        \caption{DD2}
        \label{fig:eos_panel_dd2}
    \end{subfigure}\hfill
    \begin{subfigure}[t]{0.32\textwidth}
        \centering
        \includegraphics[width=\linewidth]{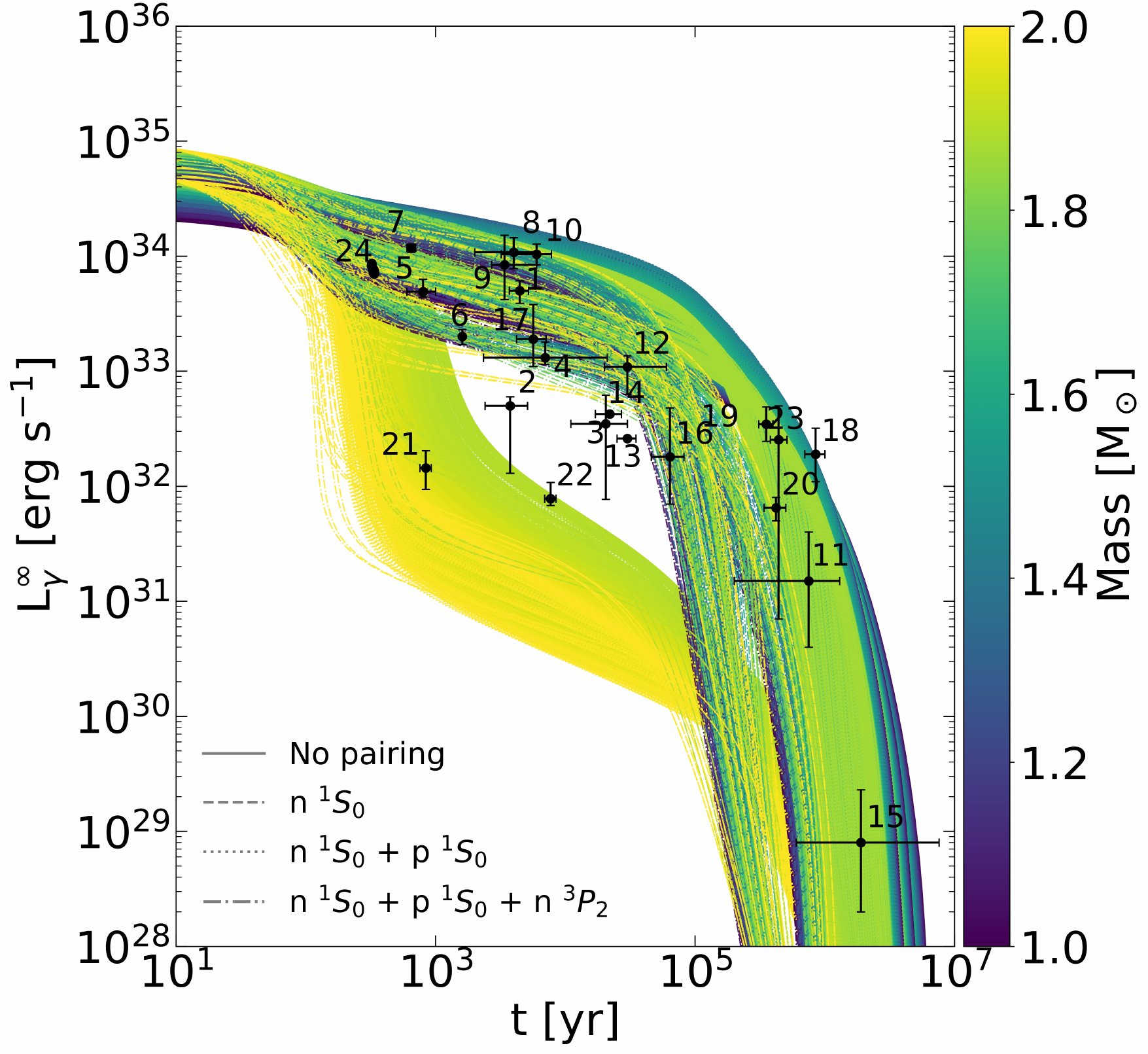}
        \caption{FSU2R}
        \label{fig:eos_panel_fsu2r}
    \end{subfigure}
    \vspace{0.35em}
    \begin{subfigure}[t]{0.32\textwidth}
        \centering
        \includegraphics[width=\linewidth]{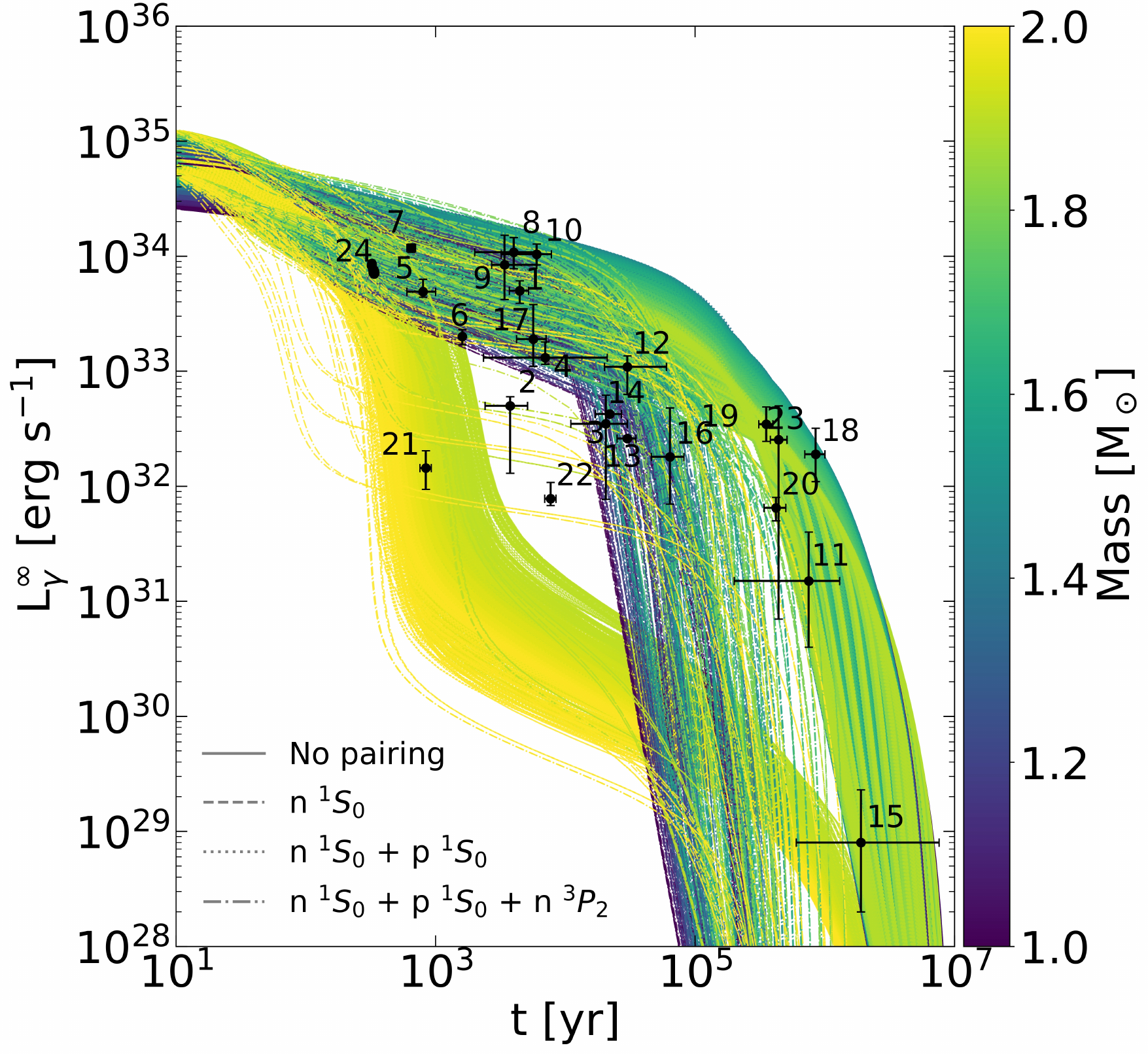}
        \caption{IST}
        \label{fig:eos_panel_ist}
    \end{subfigure}
    \hspace{0.05\textwidth}
    \begin{subfigure}[t]{0.32\textwidth}
        \centering
        \includegraphics[width=\linewidth]{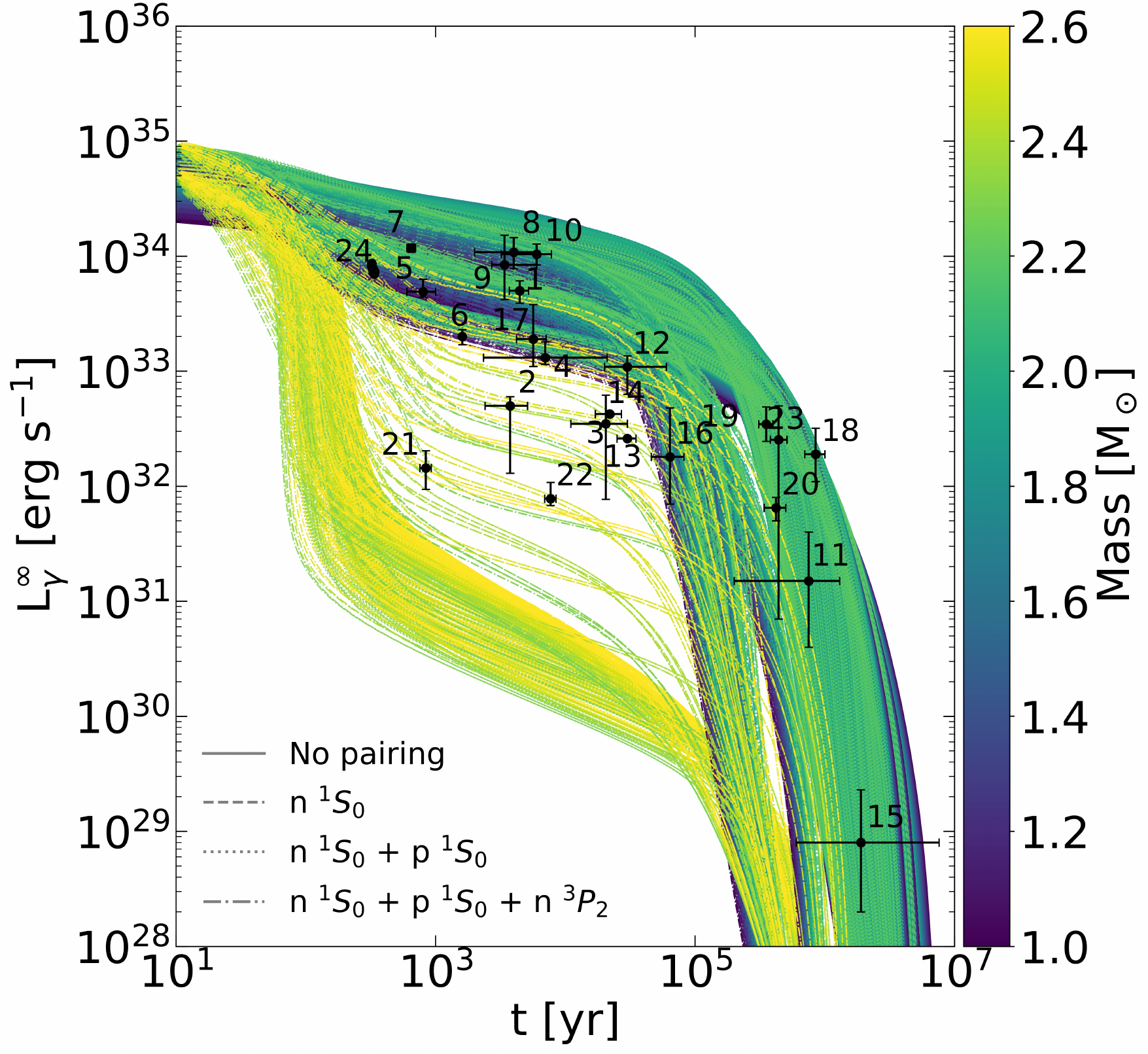}
        \caption{NL3$\omega\rho$}
        \label{fig:eos_panel_nl3wr}
    \end{subfigure}
    \caption{Cooling curves for stars with gravitational masses spanning the range indicated by the colour bar and computed using the APR4 (a), DD2 (b), FSU2R (c), IST (d), and NL3$\omega\rho$ (e) EoSs. Each panel displays the red-shifted photon luminosity, $L_\gamma^\infty$, as a function of the stellar age, $t$, for different gravitational masses, envelope compositions, and pairing combinations.The cooling curves are shown for unpaired matter (solid), models including only neutron ${}^1S_0$ pairing (dashed), both neutron and proton singlet pairing (dotted), and, for the latter case, additional neutron ${}^3P_2$ triplet pairing (dash-dotted). The utilized observational data are listed in the Appendix~\ref{app:obs_app}.}
    \label{fig:cooling_panel_all}
\end{figure*}

\subsection{Cooling processes}
The thermal evolution of an NS is described by the thermal balance equation, which relates the redshifted temperature of matter at the bottom of the envelope, $T_b^\infty$, to the redshifted neutrino and photon luminosities
\begin{equation}
C_v \frac{dT_{b}^\infty}{dt}
 = - L_\nu^\infty - L_\gamma^\infty \pm H^\infty,
\label{eq:balance}
\end{equation} 
where $t$ is the time measured by a distant observer, $C_v$ is the total heat capacity, and $T_b$ is the local matter temperature at the bottom of the envelope, typically defined at a density $\rho_b \sim 10^{10}~\mathrm{g\ cm^{-3}}$. The local effective surface temperature, $T_s$, determines the photon luminosity~\citep{Page:1997mj}. Their redshifted counterparts are $T_b^\infty$ and $T_s^\infty$, respectively. The photon luminosity measured by a distant observer is given by $L_\gamma^\infty = 4\pi R^2(T_s^\infty)^4$, where $R$ is the circumferential radius of the star. In contrast, neutrino emission occurs from the bulk of the star. The term $H^\infty$ represents additional contributions from internal heating or cooling mechanisms, e.g., magnetic field decay, dark matter effect, etc.~\citep{Ho:2014pta,Sagun:2023rzp,Avila:2023rzj,Giangrandi:2024qdb}. In this work, no additional heating sources are considered, $H^\infty = 0$.

NS cooling could be roughly split into three stages: (i)
unequilibrated stage~\citep{Sales:2020aad}; (ii) neutrino emission dominated stage, and (iii) photon emission dominated stage~\citep{Page:2004fy,Yakovlev:2004iq}. The early time up to the moment when the core and crust become thermally coupled is called the thermal relaxation time, which takes around $\sim100$–$300$ years since the formation of the NS. During this stage, the surface temperature is largely independent of the stellar mass, core neutrino luminosity, and internal composition, being mainly determined by the crustal thermal energy content, neutrino losses, and heat transport.  Once the crust and core thermally couple, the thermal evolution of the core is transmitted to the surface and becomes observable.

After thermal relaxation, the high thermal conductivity maintains an approximately uniform temperature throughout the interior below the envelope. 
The neutrino-dominated cooling stage lasts for approximately $10^5$ years and transitions to a photon-dominated phase for older NSs, producing a characteristic `shoulder'-type behaviour in the cooling curve~\citep{Page:2005fq}. The transition between these regimes depends on the particle composition of the NS and star mass. Neutrino emission in the core arises from weak interactions of degenerate baryons and leptons. The most efficient mechanisms are the $\beta$-decay
\begin{equation}
    n \rightarrow p + l^- + \overline{\nu}_{l},
\end{equation}
and its inverse 
\begin{equation}
    p + l^- \rightarrow n + {\nu}_{l},
\end{equation}
where $l$ stands for leptons, either electrons or muons. These processes require the Fermi momenta ($p_F$) of the particle species involved to satisfy a triangle inequality $p_{F,p}+p_{F,l}\geq p_{F,n}$~\citep{Lattimer:1991ib,Page:2005fq,Potekhin:2015qsa}. This condition imposes a threshold on the proton fraction, typically $\mathrm{Y_p} \gtrsim 0.11$~\citep{Lattimer:1991ib} for neutron-proton-electron ($npe$) matter and a larger value given by
\begin{equation}
\mathrm{Y_p}=\left(1+\left(1+x_e^{1/3}\right)^3 \right)^{-1},
\end{equation}
when muons are present. Here $x_e=n_e/(n_e+n_\mu)$ and the densities of electrons and muons are $n_e$ and $n_\mu$, respectively. Below this limit, only the less efficient MU, where an extra spectator nucleon enables the reaction, along with nucleon-nucleon bremsstrahlung processes, operate~\citep{Page:2004fy,Page:2005fq}.

The surface temperature of an NS is set by heat transport through its thin outer envelope, described by the $T_b$–$T_s$ relation that connects the isothermal interior to the photosphere. 
This relation is governed by a thin layer~\citep{Gudmundsson82,Gudmundsson83}, where electrons are only partially degenerate and the thermal conductivity is least efficient. Small variations in the envelope composition can significantly modify the $T_b$–$T_s$ relation by altering the thermal insulation of the star. In particular, light-element envelopes, H or He, provide a higher thermal conductivity than heavy-element ones, leading to higher surface temperatures for a given $T_b$ and resulting in systematic variations in the predicted cooling curves~\citep{Page:1997mj,Chabrier_1997,Potekhin:1997mn,Beznogov:2021ijc}. The envelope composition is parameterised by $\eta=\Delta M/M$, where $\Delta M$ denotes the mass of light elements. Typical values span $\eta\sim10^{-18}$ for almost pure heavy-element envelopes up to $\sim 10^{-6}$ for accreted light-element envelopes. In this way, the envelope provides the bridge between the NS particle composition and the X-ray observations~\citep{Potekhin:1997mn,Page:1997mj}.

\subsection{Cooling Curves}
We study the thermal evolution for different values of the NS gravitational mass, envelope composition, EoS and superconductivity/superfluidity model using NScool standard settings, not considering the effects of magnetic fields, additional cooling and/or heating mechanisms and accretion~\citep{Neutronstarcoolingcode}. 

For configurations with unpaired matter and singlet pairing, the mass grid was sampled from 1.00 to 2.00 $\mathrm{M_\odot}$ with a step of 0.01 $\mathrm{M_\odot}$ (and 0.10 $\mathrm{M_\odot}$ for triplet pairing). This upper limit was adopted for the APR4, FSU2R, and IST EoSs as the DU process is activated at gravitational masses below or close to 2 $\mathrm{M_\odot}$ (see Table~\ref{tab:astro_params}). Once DU operates, the enhanced neutrino emission leads to rapid cooling, producing a noticeable decrease in the surface luminosity within the time range covered by the observational data. For the DD2 and NL3$\omega\rho$ EoSs, the mass grid was extended beyond 2.00 $\mathrm{M_\odot}$ up to values close to the maximum gravitational mass allowed by each EoS, using a step of 0.10 $\mathrm{M_\odot}$. On the other hand, the envelope composition was parametrized by $\eta = 10^{-x}$, with $x$ ranging from 6 to 18 and sampled in steps of 1 (and 2 in the case of triplets). 

The resulting datasets contain approximately 17660 cooling curves for the APR4, FSU2R, and IST EoSs, 18480 for the DD2 EoS, and 18880 for the NL3$\omega\rho$ Eos. The resulting cooling curves are shown in Fig.~\ref{fig:cooling_panel_all}.

\section{Data Analysis}
\label{sec:stats}
The goal of this section is to confront the cooling curves generated with NSCool for different EoS, with the observed NS data shown in Fig.~\ref{fig:cooling_panel_all} and listed in Appendix~\ref{app:obs_app}. For each NS ${\nu}$, the predicted luminosity is determined jointly by the underlying EoS, the specific superconductivity and superfluidity model, and a set of nuisance parameters $\boldsymbol{\theta}^{\nu} = \left\{M_{\mathrm{NS}}^{\nu},\, \Delta M^{\nu}\right\}$, where $M_{\mathrm{NS}}^{\nu}$ is the gravitational mass of the NS, $\Delta M^{\nu}$ accounts for the envelope. We therefore consider a given model to be described by an EoS and a superconductivity/superfluidity models, labeled as `$i$-$j$-$k$', where $i$, $j$, and $k$ denote the numerical identifiers of the neutron and proton pairing gaps in the n $^1S_0$ (singlet), n $^3P_2$ (triplet), and p $^1S_0$ (singlet) channels, respectively, as listed in Table~\ref{tab:ourpairings} and described in Sec.~\ref{sec:super}. We also consider the case of unpaired matter that corresponds to `$0$-$0$-$0$'. In total, this results in $75$ different models ($5$ EoS times $15$ different pairing combinations).

Our statistical analysis follows the frequentist framework adopted in~\cite{ParticleDataGroup:2018ovx,Buschmann:2021juv}. Under the assumption that the luminosity and age measurements are independently Gaussian distributed, the likelihood for each star $\nu$ can be written as
\begin{align}
\mathcal{L}_{\nu}(\boldsymbol{d}_{\nu}| \mathrm{EoS}, \boldsymbol{\theta}^{\nu})
= & \; \mathcal{N}\left(L(\mathrm{EoS}, \boldsymbol{\theta}^{\nu}) - L_0^{\nu}, \sigma_L^{\nu}\right),
\label{eq:stats_like}
\end{align}
where the observational data set for each star ${\nu}$ is given by $\boldsymbol{d}_{\nu} = \left\{L_0^{\nu},\, \sigma_L^{\nu}\right\}$, with $L_0^{\nu}$ denoting the measured luminosity and $\sigma_L^{\nu}$ the associated uncertainty. Here, $\mathcal{N}(x,\sigma)$ denotes a normal distribution with zero mean and standard deviation $\sigma$, evaluated at $x$. In practice, we evaluate the likelihood along the theoretical cooling curve at the central value and therefore neglect time uncertainties in the analysis.
The joint-likelihood over the whole observational data is then constructed as the product over individual stars as

\begin{equation}
    \mathcal{L}(\boldsymbol{d} \mid \mathrm{EoS}, \boldsymbol{\theta}) = \prod_{{\nu}=1}^{N} \mathcal{L}_{\nu}(\boldsymbol{d}_{\nu} \mid \mathrm{EoS}, \boldsymbol{\theta}^{\nu}),
    \label{eq:9}
\end{equation}
where $N$ corresponds to the observational data. Note that the total list of model nuisance parameters is denoted by $\boldsymbol{\theta}=\{\boldsymbol{\theta^{\nu}}\}_{{\nu}=1}^{24}$. For a given EoS and superfluidity/superconductivity models, the best-fit nuisance parameters, $\hat{\boldsymbol{\theta}}$, associated with a particular observational data point are obtained by maximizing the joint likelihood~\citep{PValuesNuisance}. Under the assumption of Gaussian distributed measurements, constant variance, and zero mean, this procedure is equivalent to minimizing the corresponding chi-square
\begin{equation}
\chi^2 = \frac{\left(L(\mathrm{EoS},\boldsymbol{\theta})-L_0\right)^2}{\sigma_L^2},\label{eq:10}
\end{equation}
between the ensemble of cooling curves and the observational point~\citep{ParticleDataGroup:2018ovx}. For each source, we therefore determine the configuration of nuisance parameters that minimizes $\chi^2$, which we denote as $\chi^2_{\min}$. 

\begin{figure*}
\centering
\includegraphics[width=\textwidth]{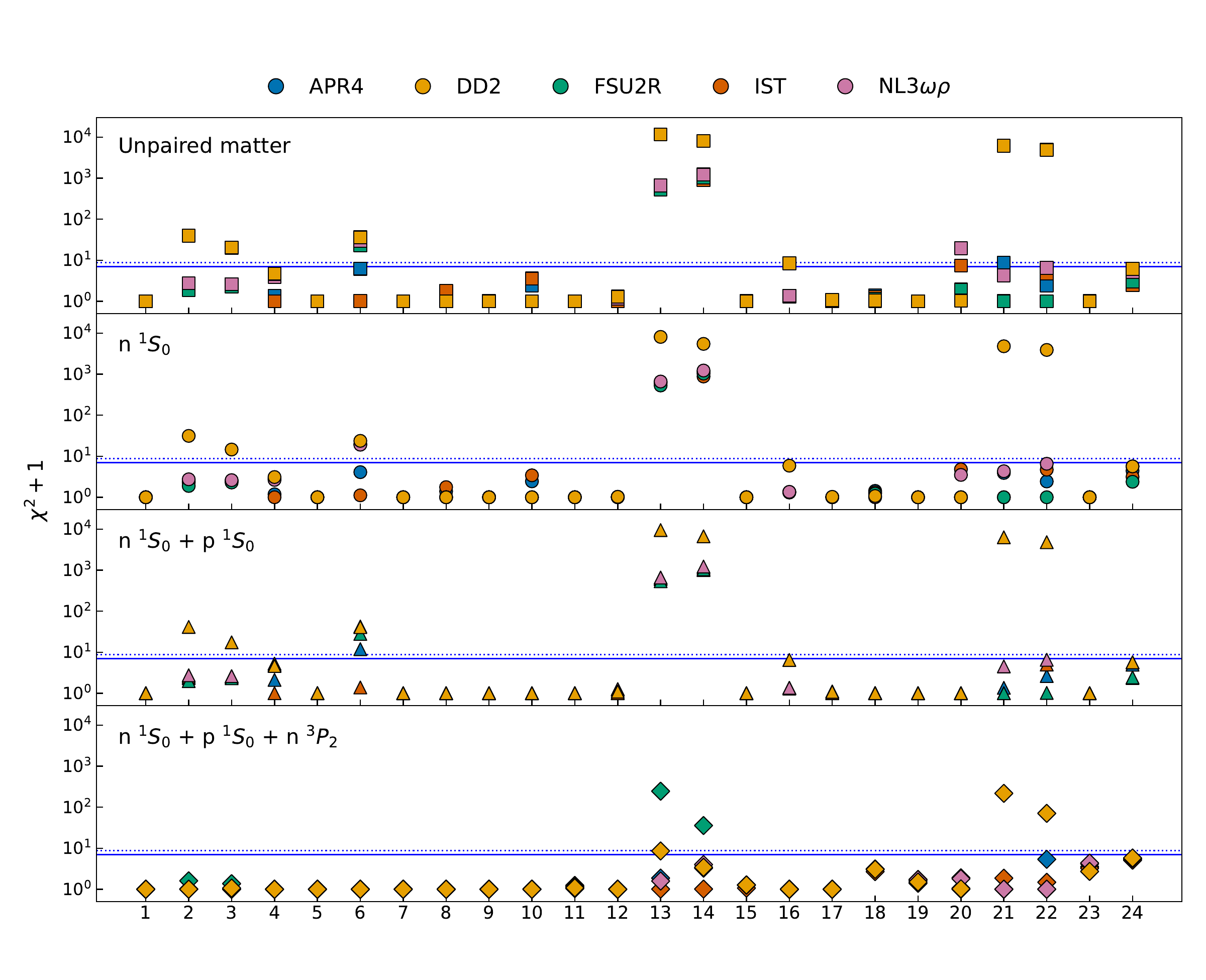}
\caption{Minimum $\chi^2$+1 values obtained for each EoS by fitting the cooling-curve dataset to each observational data point considered (see Appendix~\ref{app:obs_app} for details) for different pairing assumptions: unpaired matter (top panel), neutron $^1S_0$ pairing only (2nd panel from above), neutron and proton $^1S_0$ pairing (3rd panel from above) and including neutron $^3P_2$ pairing (bottom panel). The solid and dashed horizontal lines indicate the $2\sigma$ and $3\sigma$ confidence thresholds, respectively, drawn from a $\chi$-square distribution.}
\label{fig:chi2_all}
\end{figure*}

The analysis is carried out in two steps. First, for each observational NS data point and each EoS and pairing combination, we identify the best fitting parameters by maximizing the individual likelihood with respect to the NS mass and envelope composition. The corresponding goodness-of-fit for each NS is shown in Fig.~\ref{fig:chi2_all} for all EoSs and pairing combinations considered in this work.

For comparison we also show in the horizontal $\sigma$-lines in Fig.~\ref{fig:chi2_all} the nominal confidence levels derived from the cumulative $\chi^2$ distribution (that has been assumed). For a given number of degrees of freedom, the cumulative distribution determines the $\chi^2$ value above which the agreement between model and data is rejected at the corresponding confidence level~\citep{Cowan:2010js}. In particular, the 1, 2, and 3$\sigma$ lines correspond to confidence intervals of 68, 95, and $99.7$, respectively, obtained from the $p$-value of the cumulative $\chi^2$ distribution for the corresponding number of dof. 
Throughout this work, quoted $n\sigma$ thresholds refer to the frequentist confidence regions~\citep{ParticleDataGroup:2018ovx}.

In the second step, we require a single EoS and pairing model to describe the entire observational dataset. Using the joint likelihood in Eq.~\eqref{eq:9} and the individual chi-square contributions defined in Eq.~\eqref{eq:10}, we minimize each source's contribution over its gravitational mass and envelope composition, and sum these minima to obtain a global discrepancy statistic. We assume the resulting statistics to follow a chi-square distribution with $N-2$ degrees of freedom, $\chi_{\rm red}^2=\chi^2/\rm dof$, where $N$ is the number of luminosity measurements included in the analysis and the subtraction of two is motivated heuristically by the two stellar quantities varied for each source, $M_{\rm NS}$ and $\Delta M$. A dedicated determination of the probability distribution is needed for a more accurate determination of the exclusion lines. Therefore, the resulting thresholds are only illustrative.
Finally, we also note that we performed two separate analyses, one with all the NS cooling data, including  Cas~A~\citep{Shternin:2022rti} and another without it (see Fig. ~\ref{fig:global_chi2_plot}). 

For the observational data points used in the analysis, listed in Appendix~\ref{app:obs_app}, asymmetric uncertainties in age and luminosity are treated conservatively by adopting the larger uncertainty with respect to the central value. We adopt the luminosity uncertainties quoted in the observational compilation without adding further systematic uncertainties.

\section{Results}
\label{sec:Results}
We now discuss the results of the analysis presented in the previous section. We begin by examining the best fits for individual NSs and the role of the different superfluid and superconducting channels. This provides insight into which data points are more difficult to reproduce within the model.

We then combine these results to perform a global analysis of the full observational dataset, comparing the different EoS--pairing models using the normalized global minimum chi-square. In this global analysis, we consider two cases: one including all sources and another excluding Cas~A, in order to assess the impact of this distinctive source on the overall results.

\subsection{Individual NS fits}
We begin by fitting each NS cooling data point independently. The minimum values of $\chi^2$, obtained by varying the NS mass, the envelope composition, and the pairing models listed in Table~\ref{tab:ourpairings}, are shown in Fig.~\ref{fig:chi2_all}. The data points are labelled from 1 to 24, and different colours correspond to different EoSs. The four panels represent the various pairing scenarios: unpaired matter (top), neutron singlet pairing (second from above), neutron and proton singlet pairing (third from above), and neutron and proton singlet pairing combined with neutron triplet pairing (bottom). The solid and dotted blue lines indicate the $2\sigma$ and $3\sigma$ nominal confidence intervals, respectively.

Cooling curves from unpaired matter and singlet pairing show similar performance in reproducing the cooling data across all EoSs. Among them, the DD2 EoS exhibits some tension with the data, with seven data points showing a less satisfactory agreement. For the remaining EoSs, the most challenging cases correspond to data points 6, 13, and 14, whose $\chi^2$ values typically lie well above the $3\sigma$ threshold.

The situation changes significantly when neutron triplet pairing is included. Several data points that were previously disfavored at more than the nominal $3\sigma$ level are shifted below the $2\sigma$ confidence threshold, meaning that most data points can be reproduced for all EoS by the cooling curve of at least one combination of the models listed in Table~\ref{tab:ourpairings}. The main exceptions are the data points 13 and 14 for the FSU2R EoS, and 21 and 22 for the DD2 EoS.
\begin{figure*}
    \centering
    \includegraphics[width=1\textwidth]{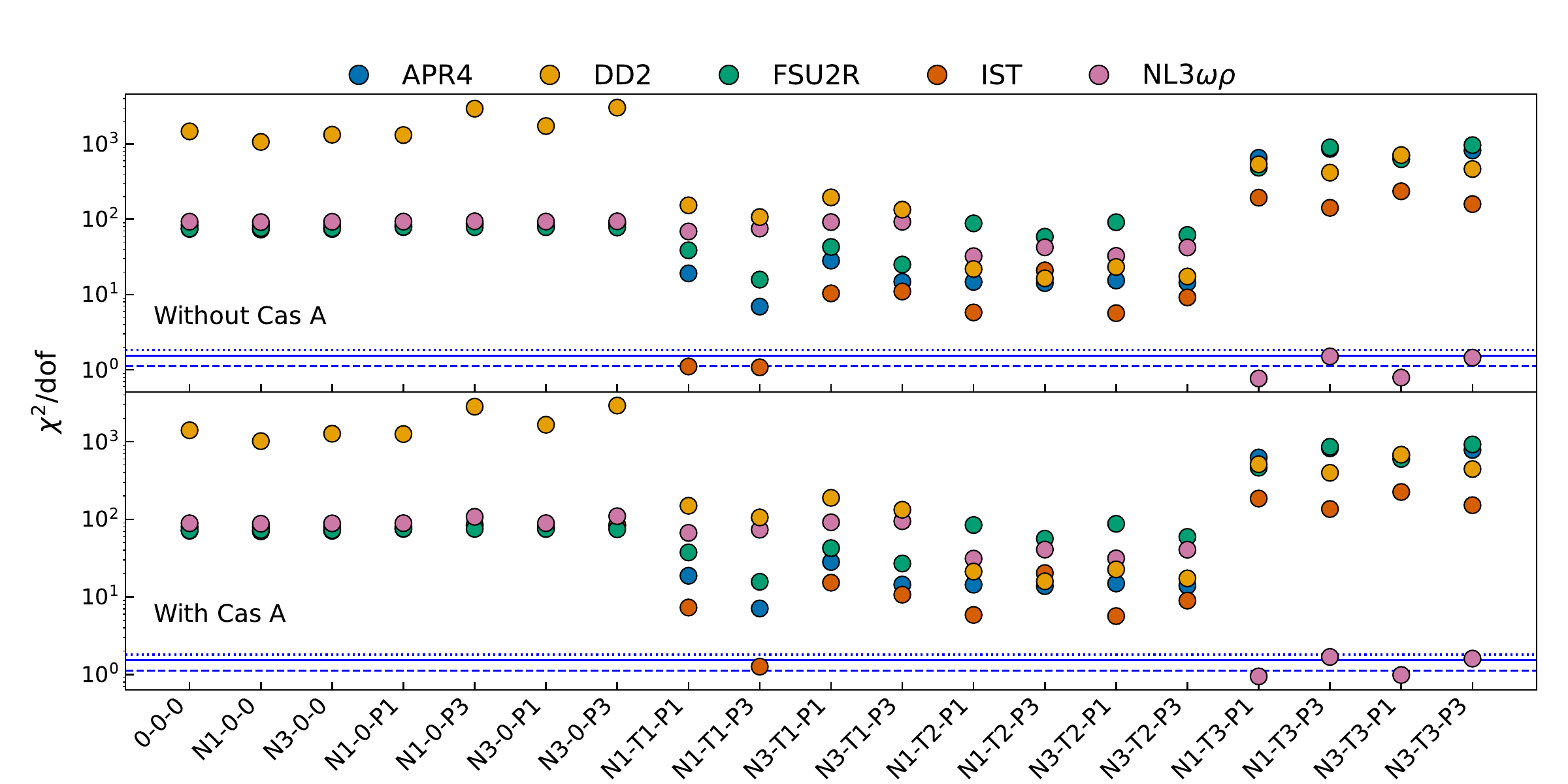}
    \caption{Reduced global $\chi^2$ over a given pairing combination considered in this work. The dashed, solid, and dotted horizontal lines indicate the $1\sigma$, $2\sigma$, and $3\sigma$ nominal confidence thresholds, respectively. The top panel shows the analysis excluding the Cas~A data, while the bottom panel includes Cas~A using the GRADED-mode observations with variable hydrogen column density (see text for details).}
    \label{fig:global_chi2_plot}
\end{figure*}

\subsection{Global minimum \texorpdfstring{$\chi^2$}{χ²}}
\label{subsec:global}
We now perform a global analysis by combining all data points and requiring each individual model—defined by a given EoS and pairing set ($i$–$j$–$k$)—to simultaneously reproduce the full cooling dataset. This is done by varying the two nuisance parameters, $M_{\mathrm{NS}}^{\nu}$ and $\Delta M^{\nu}$. The resulting reduced $\chi^2$ values for each model combination are shown in Fig.~\ref{fig:global_chi2_plot}, both excluding (upper panel) and including (lower panel) the Cas~A data (see Fig.~\ref{fig:casA}).

Overall, the vast majority of the sampled model combinations
provide a poor simultaneous description of the adopted cooling
data, as indicated by their large minimum global chi-square values.
The only clear exceptions are the NL3$\omega\rho$ models with `T3' triplet pairing and the IST model with `N1' singlet and 'T1' triplet neutron pairing. The origin of the poor fits in the remaining models can be traced back to Fig.~\ref{fig:chi2_all}, in particular to stars number 13, CXOU J061705.3+222127 (IC 443), and 14, PSR B0833-45 (Vela). These objects, characterized by very small error bars, lie in an intermediate region of the luminosity plot in Fig.~\ref{fig:cooling_panel_all}, roughly spanning $1$–$100$ kyr and $10^{31}$–$10^{33}$ erg s$^{-1}$. Within the set of EoS models considered here, this region is difficult to reproduce: models without DU processes tend to predict higher luminosities, whereas those with an active DU channel typically yield lower luminosities. Only the NL3$\omega\rho$ model with triplet pairing fully covers this intermediate regime, with the IST EoS providing a partial match. This ultimately explains why these models are the only ones capable of satisfying all observational constraints. 

In~\cite{Negreiros:2018cho}, it was shown that the FSU2R EoS can provide a satisfactory description of relatively cool stars when rapid DU cooling is combined with neutron and proton singlet pairing. However, the proton pairing adopted in that study is substantially stronger, with a critical temperature well above those considered in the present study and in the literature.

These results differ from those reported in~\cite{Das:2024ams}, where it was found that singlet pairing scenarios alone can account for the cooling data. This difference is attributed to the distinct microphysical inputs adopted in the study. In particular, Ref.~\cite{Das:2024ams} considers a relatively limited set of EoSs, among which there is the Brueckner–Hartree–Fock EoS with the Argonne V18 nucleon–nucleon potential~\citep{Baldo:2011gz} permitting a very early onset of the DU processes. In addition, the singlet pairing gaps adopted in that study are substantially larger than their typical maximum values reported in the literature and extend over a considerably broader density range, reaching densities several times higher than those commonly considered. These assumptions lead to a strong suppression of the neutrino emissivity, enabling the observational data to be reproduced. Given these differences in setup, a direct quantitative comparison between the two analyses is not straightforward.

\subsection{Cassiopeia A}
\label{subsec:cassa}
\begin{figure}
\centering
\setkeys{Gin}{width=0.95\columnwidth}
\includegraphics{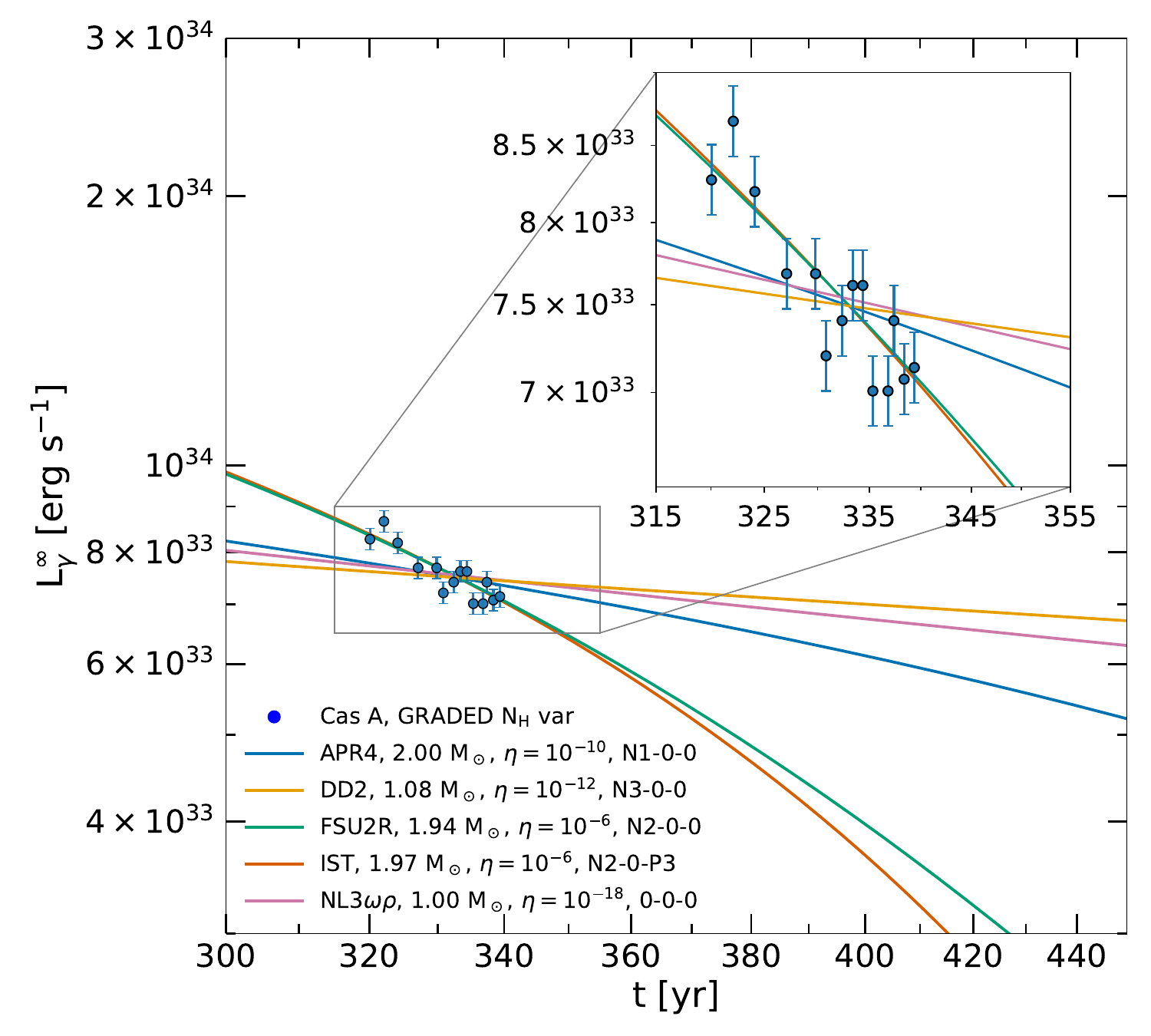}
\caption{Red-shifted photon luminosity, $L_\gamma^\infty$, as a function of stellar age, $t$, for the best-fit cooling curves corresponding to the considered EoSs - APR4, DD2, FSU2R, IST, and NL3$\omega\rho$. The observational data corresponds to the data presented in Table~\ref{tab:casa_graded_var_data}.}
\label{fig:casA}
\end{figure}

We now briefly discuss Cas~A, which is distinctive within the dataset containing two decades of observation (Fig.~\ref{fig:casA}). For Cas~A, we used the observations obtained by \textit{Chandra} ACIS in GRADED mode, adopting the data in which the hydrogen column density, $N_{\mathrm H}$, is treated as a free parameter~\citep{Shternin:2022rti} (see Table~\ref{tab:casa_graded_var_data} in Appendix). 

We find that including Cas~A does not drastically modify the overall global fit, since the reduced $\chi^2$ values change only moderately across the EoSs and pairing configurations considered. However, it is worth noting that Cas~A introduces a preference in the parameter space. Fig.~\ref{fig:casA} shows the best fits to the Cas~A data. The FSU2R EoS accompanied with singlet neutron pairing and the IST EoS including singlet neutron and strong proton pairing provide the best agreement. The corresponding curves are obtained for heavy NSs with $M \gtrsim 1.94$--$1.97 \mathrm{M_\odot}$.

Although Cas~A has a limited impact on the overall global fit, it can substantially affect the statistical viability of specific pairing configurations. In particular, the inclusion of Cas~A shifts the fixed pairing combination N1-T1-P1 for the IST EoS from being compatible at the nominal $2\sigma$ level to exceeding the $3\sigma$ confidence threshold. Similar results were obtained in~\cite{Tsiopelas:2020nzm,Tsiopelas:2021ayd} for the IST EoS. This shows that, despite its modest statistical weight within the full sample, Cas~A acts as a strong discriminator for particular pairing combinations within otherwise acceptable cooling scenarios. This is presented in the bottom panel of Fig.~\ref{fig:global_chi2_plot}.
%
\section{Conclusions and discussion}
\label{sec:Conclusion}
We presented a frequentist $\chi^2$ analysis of the thermal evolution of isolated NSs by confronting a large grid of cooling curves obtained with \textsc{NSCool} code with the luminosity-age observations of 24 sources. For each object, we varied the gravitational mass and envelope composition and explored five nucleonic EoSs: APR4, DD2, FSU2R, IST, and NL3$\omega\rho$, combined with and without representative proton singlet, neutron singlet and triplet pairing models.

We found that models without pairing provide a poor description of the data. Including neutron and proton singlet pairing improves the agreement for a subset of sources, but remains overall insufficient, leaving several objects in strong tension with the data. Particularly, four middle-age objects with a lower surface temperature challenge the NS thermal evolution modeling: CXOU J061705.3+222127 (IC 443) (number 13), PSR B0833-45 (Vela) (number 14), PSR J0205+6449 (3C 58) (number 21), and PSR B2334+61 (number 22). 

Within the set of EoSs and superfluidity/superconductivity models considered, a consistent description of the full dataset is only achieved when neutron triplet pairing is included; even in this case, viable fits are restricted to the NL3$\omega\rho$ EoS and, to a lesser extent, the IST EoS. The remaining EoSs explored, namely, the APR4, DD2, and FSU2R, are therefore disfavoured. We also assessed the impact of the Cas~A data on the overall analysis and found that they do not qualitatively alter the global fit values across the different EoSs, although they can discriminate between specific fixed pairing configurations.

The two cooler middle-age objects, PSR J0205+6449 (3C 58) and PSR B2334+61, together with CXOU J085201.4-461753 (Vela Jr.), were also identified in~\cite{Marino:2024gpm} as objects that cannot be explained without enhanced cooling. In~\cite{Marino:2024gpm}, the authors found that the DU process needs to be activated to reproduce the observed surface temperatures of these objects. Providing an efficient neutrino-emission channel capable of explaining faintest middle-age NSs. However, we find that models with active DU cooling tend to over-suppress the luminosity at later times, making it difficult to reproduce the data in the central region of the luminosity-age diagram. This tension is alleviated when neutron triplet pairing is included.

The need for enhanced cooling through the DU process has also been reported in~\citep{Beznogov:2014yia,Beznogov:2015ewa,Das:2024ams}. These studies showed that singlet proton pairing can efficiently suppress the neutrino emission, while the PBF process associated with neutron ${}^3P_2$ pairing does not always provide a satisfactory description of the data. While the results regarding the active cooling channels and pairing gaps are not yet conclusive, recent studies broadly indicate that enhanced neutrino cooling is required to reproduce the observational data, particularly for Cas~A and the cooler middle-aged objects. The results suggest that activated DU cooling can be too strong and therefore requires a compensating effect, either through singlet proton or neutron triplet pairing. Alternatively, rapid cooling could be explained by in-medium enhancements of the MU neutrino reaction rates, as investigated in~\citep{Grigorian:2005fn,Blaschke:2013vma,Shternin:2018dcn}. This problem is intrinsically multiparametric, as the choice of the underlying EoS, and hence the proton fraction and symmetry energy, affects the required strength of the pairing gaps and pairing channels. These results demonstrate that, in addition to multimessenger observations, NS cooling provides complementary constraints on dense matter, placing important constraints on the proton fraction and symmetry energy of realistic nucleonic NS EoSs alongside those obtained from nuclear structure, global NS observables, and recent gravitational-wave observations, while offering new insights into the interplay between the EoS and nucleon pairing properties.

As a further extension of the present study, it will be important to broaden the range of EoSs considered in the analysis to assess whether the apparent need for neutron triplet pairing found here is a generic feature across different EoSs or instead specific to the subset explored in this study. Such an extension could provide further insight into the underlying nuclear physics. Another natural extension would be to incorporate independent constraints on NS masses. Our current analysis does not make use of mass measurements, which in some cases could provide significant additional constraining power. This may be particularly relevant for well-studied objects such as Cas~A. Finally, it would be interesting to investigate the impact of magnetic field and assess how its inclusion may affect the conclusions drawn here.

\section*{Acknowledgements}
A.A. and R.Z.F acknowledge the financial support provided by FCT - Fundação para a Ciência e Tecnologia, I.P., through  
the Strategic Funding UID/04650/2025 and UID/04564/2025 and 
national funds with DOI identifiers 10.54499/2023.11681.PEX, 10.54499/2024.00249.CERN funded by measure RE-C06-i06.m02 – ``Reinforcement of funding for International Partnerships in Science, Technology and Innovation'' of the Recovery and Resilience Plan - RRP, within the framework of the financing contract signed between the Recover Portugal Mission Structure (EMRP) and the FCT - Fundação para a Ciência e Tecnologia, I.P. as an intermediate beneficiary. A.A. also acknowledge the support by FCT - Fundação para a Ciência e Tecnologia, I.P. by project reference 2025.01054.BD and DOI identifier 10.54499/2025.01054.BD. 
V.S. gratefully acknowledges support from the UKRI-funded ``The next-generation gravitational-wave observatory network'' project (Grant No. ST/Y004248/1). V.S. acknowledges the gratitude to the FCT for support under Advanced Computing Projects 2024.07037.CPCA.A1 with DOI identifier 10.54499/2024.07037.CPCA.A1. The authors also acknowledge the support from the FCT within the project No. 2023.10526.CPCA.A2 with DOI identifier 10.54499/2023.10526.CPCA.A2.

\section*{Data availability}
The cooling data underlying this article are publicly available via the Zenodo database at 10.5281/zenodo.22102266.

\appendix
\section{Observational data}
\label{app:obs_app}

\begin{table*}
\centering
\setlength{\tabcolsep}{4pt}
\renewcommand{\arraystretch}{1.25}
\begin{tabular}{clccc}
\hline
No. & Object name & $t$ [kyr] & $L_\gamma^\infty$ [$10^{33}$ erg s$^{-1}$] & Refs. \\
\hline
1 & RX J0822.0-4300 (Puppis A) & $4.45^{+0.75}_{-0.75}$ & $5^{+1.1}_{-1.1}$ &~\citep{Becker:2012dj,DeLuca:2011yr} \\
2 & CXOU J085201.4-461753 (Vela Jr.) & $3.075^{+2.025}_{-0.675}$ & $0.5^{+0.1}_{-0.37}$ &~\citep{Allen:2014yra,Ho:2026guu,Marino:2024gpm} \\
3 & 2XMM J104608.7-594306 (Homunculus) & $20.5^{+9.5}_{-9.5}$ & $0.348^{+0.272}_{-0.271}$ &~\citep{Pires:2015gna} \\
4 & 1E 1207.4-5209 (G296.5+10.0) & $7^{+14}_{-4.667}$ & $1.31^{+0.49}_{-0.16}$ &~\citep{Roger:1988,Mereghetti:2002za} \\
5 & CXOU J160103.1-513353 (G330.2+1.0) & $0.8^{+0.2}_{-0.2}$ & $4.9^{+1.4}_{-0.5}$ &~\citep{Borkowski:2018omn,Ho:2021hwy} \\
6 & 1WGA J1713.4-3949 (G347.3-0.5) & $1.6085^{+0.0005}_{-0.0005}$ & $2^{+0.3}_{-0.3}$ &~\citep{Ho:2021hwy,Cassam-Chenai:2004jon} \\
7 & XMMU J172054.5-372652 (G350.1-0.3) & $0.65^{+0.05}_{-0.05}$ & $11.8^{+1}_{-1}$ &~\citep{Mayer:2021wvs,Ho:2021hwy} \\
8 & XMMU J173203.3-344518 (HESS J1731-347) & $4^{+2}_{-2}$ & $10.8^{+3.7}_{-3}$ &~\citep{Maxted:2017xfs,Doroshenko:2022nwp}\\
9 & CXOU J181852.0-150213 (G015.9+00.2) & $3.4^{+2.6}_{-0.7}$ & $8.4^{+6.8}_{-4.2}$ &~\citep{Sasaki:2018wnz,Klochkov:2016lly} \\
10 & CXOU J185238.6+004020 (Kes 79) & $6^{+1.8}_{-2.8}$ & $10.4^{+2.4}_{-2}$ &~\citep{Sun:2004hz,Bogdanov:2014hna} \\
11 & PSR J0357+3205 (Morla) & $750^{+550}_{-550}$ & $0.015^{+0.025}_{-0.011}$ &~\citep{Kirichenko:2014ona} \\
12 & PSR J0538+2817 (Sim 147) & $30^{+30}_{-10}$ & $1.09^{+0.27}_{-0.46}$ &~\citep{Ng:2006vh,Kramer:2003au} \\
13 & CXOU J061705.3+222127 (IC 443) & $30^{+5}_{-5}$ & $0.26^{+0.01}_{-0.01}$ &~\citep{Chevalier:1998ew,Swartz:2015fwa} \\
14 & PSR B0833-45 (Vela) & $22^{+5}_{-5}$ & $0.424^{+0.012}_{-0.012}$ &~\citep{Aschenbach:2002es,Ofengeim:2018qkr} \\
15 & PSR B0950+08 & $1900^{+5700}_{-1300}$ & $0.00008^{+0.00015}_{-0.00006}$ &~\citep{Igoshev:2018fcp,Abramkin:2021fzy} \\
16 & PSR B1951+32 (CTB 80) & $64^{+18}_{-18}$ & $0.18^{+0.3}_{-0.11}$ &~\citep{Migliazzo:2002vd,Li:2005qa} \\
17 & PSR J1119-6127 (G292.2-00.5) & $5.65^{+1.45}_{-1.45}$ & $1.9^{+1.9}_{-0.8}$ &~\citep{Kumar:2012dm,Ng:2012dv} \\
18 & RX J0720.4-3125 & $850^{+150}_{-150}$ & $0.19^{+0.13}_{-0.08}$ &~\citep{Tetzlaff:2011kh,Hambaryan:2017wvm} \\
19 & RX J1605.3+3249 & $440^{+70}_{-60}$ & $0.253^{+0.246}_{-0.246}$ &~\citep{Tetzlaff:2012rz,Malacaria:2019zqr,Pires:2019qsk} \\
20 & RX J1856.5-3754 (Upper Scorpius) & $420^{+80}_{-80}$ & $0.065^{+0.015}_{-0.015}$ &~\citep{Mignani:2012mm,Yoneyama:2017xth} \\
21 & PSR J0205+6449 (3C 58) & $0.841^{+0.084}_{-0.084}$ & $0.144^{+0.06}_{-0.05}$ &~\citep{Kothes:2013hka,Potekhin:2020ttj,Marino:2024gpm} \\
22 & PSR B2334+61 & $7.7^{+0.77}_{-0.77}$ & $0.078^{+0.03}_{-0.01}$ &~\citep{Marino:2024gpm} \\
23 & PSR B0656+14 (Monogem Ring) & $353.18^{+43.65}_{-43.65}$ & $0.347^{+0.143}_{-0.101}$ &~\citep{Suzuki:2021ium,Zharikov:2021llh} \\
\hline
\end{tabular}
\caption[Observational dataset used in this work.]{Observational dataset used in this work, taken from~\cite{Potekhin:2020ttj} (see also~\cite{PotekhinCoolData}). Stellar age is presented in kyr and luminosities, $L_\gamma^\infty$, are given in units of $10^{33}$~$\mathrm{erg\ s^{-1}}$.}
\label{tab:observational_data}
\end{table*}
\begin{table*}
\centering
\renewcommand{\arraystretch}{1.25}
\begin{tabular}{clccc}
\hline
No. & Object name & $t$ [yr] & $L_\gamma^\infty$ [$10^{33}$ erg s$^{-1}$] & Refs. \\
\hline
\multirow{14}{*}{24} 
& \multirow{14}{*}{Cas~A (GRADED N$_H$ var)} 
& $320.08$ & $8.272^{+0.232}_{-0.225}$ & \multirow{14}{*}{\citep{Shternin:2022rti}} \\
& & $322.10$ & $8.662^{+0.243}_{-0.236}$ & \\
& & $324.10$ & $8.197^{+0.230}_{-0.223}$ & \\
& & $327.09$ & $7.685^{+0.215}_{-0.209}$ & \\
& & $329.84$ & $7.685^{+0.215}_{-0.209}$ & \\
& & $330.83$ & $7.205^{+0.202}_{-0.196}$ & \\
& & $332.37$ & $7.407^{+0.208}_{-0.202}$ & \\
& & $333.39$ & $7.614^{+0.213}_{-0.208}$ & \\
& & $334.36$ & $7.614^{+0.213}_{-0.208}$ & \\
& & $335.33$ & $7.009^{+0.196}_{-0.191}$ & \\
& & $336.80$ & $7.009^{+0.196}_{-0.191}$ & \\
& & $337.38$ & $7.407^{+0.208}_{-0.202}$ & \\
& & $338.37$ & $7.073^{+0.198}_{-0.193}$ & \\
& & $339.37$ & $7.139^{+0.199}_{-0.195}$ & \\
\hline
\end{tabular}
\caption{Cas~A observations with \textit{Chandra} ACIS in GRADED mode, whereas the hydrogen column density, $N_{\mathrm H}$, is variable. Time is given in years, and luminosities are expressed in units of $10^{33} \mathrm{erg\ s^{-1}}$.}
\label{tab:casa_graded_var_data}
\end{table*}




\bibliographystyle{mnras}
\bibliography{bibliography}

@article{Oppenheimer:1939ne,
    author = "Oppenheimer, J. R. and Volkoff, G. M.",
    title = "{On massive neutron cores}",
    doi = "10.1103/PhysRev.55.374",
    journal = "Phys. Rev.",
    volume = "55",
    pages = "374--381",
    year = "1939"
}

@article{Tolman:1939jz,
    author = "Tolman, Richard C.",
    title = "{Static solutions of Einstein's field equations for spheres of fluid}",
    doi = "10.1103/PhysRev.55.364",
    journal = "Phys. Rev.",
    volume = "55",
    pages = "364--373",
    year = "1939"
}

@inproceedings{Page:1997mj,
    author = "Page, Dany",
    title = "{Thermal evolution of isolated neutron stars}",
    booktitle = "{NATO Advanced Study Institute: The Many Faces of Neutron Stars}",
    eprint = "astro-ph/9706259",
    archivePrefix = "arXiv",
    month = "6",
    year = "1997"
}

@article{Ho:2014pta,
    author = "Ho, Wynn C. G. and Elshamouty, Khaled G. and Heinke, Craig O. and Potekhin, Alexander Y.",
    title = "{Tests of the nuclear equation of state and superfluid and superconducting gaps using the Cassiopeia A neutron star}",
    eprint = "1412.7759",
    archivePrefix = "arXiv",
    primaryClass = "astro-ph.HE",
    doi = "10.1103/PhysRevC.91.015806",
    journal = "Phys. Rev. C",
    volume = "91",
    number = "1",
    pages = "015806",
    year = "2015"
}

@article{Avila:2023rzj,
    author = "{\'A}vila, Afonso and Giangrandi, Edoardo and Sagun, Violetta and Ivanytskyi, Oleksii and Provid{\^e}ncia, Constan{\c{c}}a",
    title = "{Rapid neutron star cooling triggered by dark matter}",
    eprint = "2309.03894",
    archivePrefix = "arXiv",
    primaryClass = "astro-ph.HE",
    doi = "10.1093/mnras/stae337",
    journal = "Mon. Not. Roy. Astron. Soc.",
    volume = "528",
    number = "4",
    pages = "6319--6328",
    year = "2024"
}

@article{Giangrandi:2024qdb,
    author = "Giangrandi, Edoardo and {\'A}vila, Afonso and Sagun, Violetta and Ivanytskyi, Oleksii and Provid{\^e}ncia, Constan{\c{c}}a",
    title = "{The Impact of Asymmetric Dark Matter on the Thermal Evolution of Nucleonic and Hyperonic Compact Stars}",
    eprint = "2401.03295",
    archivePrefix = "arXiv",
    primaryClass = "astro-ph.HE",
    doi = "10.3390/particles7010010",
    journal = "Particles",
    volume = "7",
    number = "1",
    pages = "179--200",
    year = "2024"
}

@article{Sagun:2023rzp,
    author = "Sagun, Violetta and Giangrandi, Edoardo and Dietrich, Tim and Ivanytskyi, Oleksii and Negreiros, Rodrigo and Provid{\^e}ncia, Constan{\c{c}}a",
    title = "{What Is the Nature of the HESS J1731-347 Compact Object?}",
    eprint = "2306.12326",
    archivePrefix = "arXiv",
    primaryClass = "astro-ph.HE",
    doi = "10.3847/1538-4357/acfc9e",
    journal = "Astrophys. J.",
    volume = "958",
    number = "1",
    pages = "49",
    year = "2023"
}

@article{Page:2004fy,
    author = "Page, Dany and Lattimer, James M. and Prakash, Madappa and Steiner, Andrew W.",
    title = "{Minimal cooling of neutron stars: A New paradigm}",
    eprint = "astro-ph/0403657",
    archivePrefix = "arXiv",
    doi = "10.1086/424844",
    journal = "Astrophys. J. Suppl.",
    volume = "155",
    pages = "623--650",
    year = "2004"
}

@article{Yakovlev:2004iq,
    author = "Yakovlev, Dima G. and Pethick, C. J.",
    title = "{Neutron star cooling}",
    eprint = "astro-ph/0402143",
    archivePrefix = "arXiv",
    doi = "10.1146/annurev.astro.42.053102.134013",
    journal = "Ann. Rev. Astron. Astrophys.",
    volume = "42",
    pages = "169--210",
    year = "2004"
}

@article{Shternin:2021fpt,
    author = "Shternin, Peter S. and Ofengeim, Dmitry D. and Ho, Wynn C. G. and Heinke, Craig O. and Wijngaarden, M. J. P. and Patnaude, Daniel J.",
    title = "{Model-independent constraints on superfluidity from the cooling neutron star in Cassiopeia~A}",
    eprint = "2106.05692",
    archivePrefix = "arXiv",
    primaryClass = "astro-ph.HE",
    doi = "10.1093/mnras/stab1695",
    journal = "Mon. Not. Roy. Astron. Soc.",
    volume = "506",
    number = "1",
    pages = "709--726",
    year = "2021"
}

@article{Das:2024ams,
    author = "Das, H. C. and Wei, Jin-Biao and Burgio, G. F. and Schulze, H. -J.",
    title = "{Neutron star cooling and mass distributions}",
    eprint = "2403.02222",
    archivePrefix = "arXiv",
    primaryClass = "astro-ph.HE",
    doi = "10.1103/PhysRevD.109.123018",
    journal = "Phys. Rev. D",
    volume = "109",
    number = "12",
    pages = "123018",
    year = "2024"
}

@article{Beznogov:2014yia,
    author = "Beznogov, M. V. and Yakovlev, D. G.",
    title = "{Statistical theory of thermal evolution of neutron stars}",
    eprint = "1411.6803",
    archivePrefix = "arXiv",
    primaryClass = "astro-ph.SR",
    doi = "10.1093/mnras/stu2506",
    journal = "Mon. Not. Roy. Astron. Soc.",
    volume = "447",
    number = "2",
    pages = "1598--1609",
    year = "2015"
}

@article{Fortin:2021umb,
    author = "Fortin, Morgane and Raduta, Adriana R. and Avancini, Sidney and Provid{\^e}ncia, Constan{\c{c}}a",
    title = "{Thermal evolution of relativistic hyperonic compact stars with calibrated equations of state}",
    eprint = "2102.07565",
    archivePrefix = "arXiv",
    primaryClass = "nucl-th",
    doi = "10.1103/PhysRevD.103.083004",
    journal = "Phys. Rev. D",
    volume = "103",
    number = "8",
    pages = "083004",
    year = "2021"
}

@article{Lattimer:1991ib,
    author = "Lattimer, J. M. and Prakash, M. and Pethick, C. J. and Haensel, P.",
    title = "{Direct URCA process in neutron stars}",
    doi = "10.1103/PhysRevLett.66.2701",
    journal = "Phys. Rev. Lett.",
    volume = "66",
    pages = "2701--2704",
    year = "1991"
}

@article{Page:2005fq,
    author = "Page, Dany and Geppert, Ulrich and Weber, Fridolin",
    title = "{The Cooling of compact stars}",
    eprint = "astro-ph/0508056",
    archivePrefix = "arXiv",
    doi = "10.1016/j.nuclphysa.2005.09.019",
    journal = "Nucl. Phys. A",
    volume = "777",
    pages = "497--530",
    year = "2006"
}

@article{Potekhin:2015qsa,
    author = "Potekhin, A. Y. and Pons, J. A. and Page, Dany",
    title = "{Neutron stars - cooling and transport}",
    eprint = "1507.06186",
    archivePrefix = "arXiv",
    primaryClass = "astro-ph.HE",
    doi = "10.1007/s11214-015-0180-9",
    journal = "Space Sci. Rev.",
    volume = "191",
    number = "1-4",
    pages = "239--291",
    year = "2015"
}

@phdthesis{Yanagi:2019zne,
    author = "Yanagi, Keisuke",
    title = "{Thermal Evolution of Neutron Stars as a Probe of Physics beyond the Standard Model}",
    eprint = "2003.08199",
    archivePrefix = "arXiv",
    primaryClass = "hep-ph",
    school = "Tokyo U.",
    month = "12",
    year = "2019"
}

@article{Antoniadis:2013pzd,
	author = "Antoniadis, J. and Freire, P. C. C. and Wex, N. and Tauris, T. M. and Lynch, R. S. and van kerkwijk, M. H. and Kramer, M. and others",
	title = "{A massive pulsar in a compact relativistic binary}",
	eprint = "1304.6875",
	archivePrefix = "arXiv",
	primaryClass = "astro-ph.HE",
	doi = "10.1126/science.1233232",
	journal = "Science",
	volume = "340",
	pages = "6131",
	year = "2013"
}

@article{Romani:2021xmb,
	author = "Romani, R. W. and Kandel, D. and Filippenko, A. V. and Brink, T. G. and Zheng, W.",
	title = "{PSR J1810+1744: companion darkening and a precise high neutron star mass}",
	eprint = "2101.09822",
	archivePrefix = "arXiv",
	primaryClass = "astro-ph.HE",
	doi = "10.3847/2041-8213/abe2b4",
	journal = "Astrophys. J. Lett.",
	volume = "908",
	number = "2",
	pages = "L46",
	year = "2021"
}

@article{Romani:2022jhd,
	author = "Romani, R. W. and Kandel, D. and Filippenko, A. V. and Brink, T. G. and Zheng, W.",
	title = "{PSR J0952\ensuremath{-}0607: the fastest and heaviest known galactic neutron star}",
	eprint = "2207.05124",
	archivePrefix = "arXiv",
	primaryClass = "astro-ph.HE",
	doi = "10.3847/2041-8213/ac8007",
	journal = "Astrophys. J. Lett.",
	volume = "934",
	number = "2",
	pages = "L17",
	year = "2022"
}

@article{Miller:2019cac,	
	author = "Miller, M. C. and Lamb, F. K. and Dittmann, A.J. and Bogdanov, S. and Arzoumanian, Z. and Gendreau, K. C. and Guillot, S. and others",
	title = "{PSR J0030+0451 mass and radius from $NICER$ data and implications for the properties of neutron star matter}",
	eprint = "1912.05705",
	archivePrefix = "arXiv",
	primaryClass = "astro-ph.HE",
	doi = "10.3847/2041-8213/ab50c5",
	journal = "Astrophys. J. Lett.",
	volume = "887",
	number = "1",
	pages = "L24",
	year = "2019"
}

@article{LIGOScientific:2018cki,
	author = "Abbott, B. P. and Abbott, R. and Abbott. T.D. and Acernese, F. and Ackley, K. and Adams, C. and Adams, T. and others",
	collaboration = "LIGO Scientific, Virgo",
	title = "{GW170817: measurements of neutron star radii and equation of state}",
	eprint = "1805.11581",
	archivePrefix = "arXiv",
	primaryClass = "gr-qc",
	reportNumber = "LIGO-P1800115",
	doi = "10.1103/PhysRevLett.121.161101",
	journal = "Phys. Rev. Lett.",
	volume = "121",
	number = "16",
	pages = "161101",
	year = "2018"
}

@article{LIGOScientific:2020aai,
	author = "Abbott, B. P. and Abbott, R. and Abbott. T.D. and Abraham, S. and Acernese, F. and Ackley, K. and Adams, C. and others",
	collaboration = "LIGO Scientific, Virgo",
	title = "{GW190425: observation of a compact binary coalescence with total mass $\sim 3.4 M_{\odot}$}",
	eprint = "2001.01761",
	archivePrefix = "arXiv",
	primaryClass = "astro-ph.HE",
	reportNumber = "LIGO-P190425",
	doi = "10.3847/2041-8213/ab75f5",
	journal = "Astrophys. J. Lett.",
	volume = "892",
	number = "1",
	pages = "L3",
	year = "2020"
}

@article{Gudmundsson83,
author = {Gudmundsson, Einar and Pethick, C. and Epstein, Richard},
year = {1983},
month = {10},
pages = {286},
title = {Structure of neutron star envelopes},
volume = {272},
journal = {Astrophys. J.},
doi = {10.1086/161292}
}

@article{Marino:2024gpm,
    author = "Marino, Alessio and Dehman, Clara and Kovlakas, Konstantinos and Rea, Nanda and Pons, Jose A. and Vigan{\`o}, D.",
    title = "{Constraints on the dense matter equation of state from young and cold isolated neutron stars}",
    eprint = "2404.05371",
    archivePrefix = "arXiv",
    primaryClass = "astro-ph.HE",
    doi = "10.1038/s41550-024-02291-y",
    journal = "Nature Astron.",
    volume = "8",
    number = "8",
    pages = "1020--1030",
    year = "2024"
}

@article{Gudmundsson82,
author = {Gudmundsson, Einar and Pethick, C. and Epstein, Richard},
year = {1982},
month = {08},
title = {Neutron star envelopes},
volume = {259},
journal = {Astrophys. J.},
doi = {10.1086/183840}
}

@article{Grigorian:2005fn,
    author = "Grigorian, H. and Voskresensky, Dmitri N.",
    title = "{Medium effects in cooling of neutron stars and 3P(2) neutron gap}",
    eprint = "astro-ph/0507061",
    archivePrefix = "arXiv",
    doi = "10.1051/0004-6361:20052799",
    journal = "Astron. Astrophys.",
    volume = "444",
    pages = "913",
    year = "2005"
}

@article{Shternin:2018dcn,
    author = "Shternin, P. S. and Baldo, M. and Haensel, P.",
    title = "{In-medium enhancement of the modified Urca neutrino reaction rates}",
    eprint = "1807.06569",
    archivePrefix = "arXiv",
    primaryClass = "astro-ph.HE",
    doi = "10.1016/j.physletb.2018.09.035",
    journal = "Phys. Lett. B",
    volume = "786",
    pages = "28--34",
    year = "2018"
}

@article{Chabrier_1997,
doi = {10.1086/310535},
url = {https://doi.org/10.1086/310535},
year = {1997},
month = {mar},
publisher = {},
volume = {477},
number = {2},
pages = {L99},
author = {Chabrier, Gilles and Potekhin, Alexander Y. and Yakovlev, Dmitry G.},
title = {Cooling Neutron Stars with Accreted Envelopes},
journal = {Astrophys. J.},
}

@article{Potekhin:1997mn,
    author = "Potekhin, A. Y. and Chabrier, G. and Yakovlev, D. G.",
    title = "{Internal temperatures and cooling of neutron stars with accreted envelopes}",
    eprint = "astro-ph/9706148",
    archivePrefix = "arXiv",
    journal = "Astron. Astrophys.",
    volume = "323",
    pages = "415",
    year = "1997"
}

@article{Lombardo:2000ec,
    author = "Lombardo, U. and Schulze, H. J.",
    editor = "Blaschke, D. and Glendenning, N. K. and Sedrakian, A.",
    title = "{Superfluidity in neutron star matter}",
    eprint = "astro-ph/0012209",
    archivePrefix = "arXiv",
    journal = "Lect. Notes Phys.",
    volume = "578",
    pages = "30--53",
    year = "2001"
}

@article{Dean:2002zx,
    author = "Dean, D. J. and Hjorth-Jensen, M.",
    title = "{Pairing in nuclear systems: From neutron stars to finite nuclei}",
    eprint = "nucl-th/0210033",
    archivePrefix = "arXiv",
    doi = "10.1103/RevModPhys.75.607",
    journal = "Rev. Mod. Phys.",
    volume = "75",
    pages = "607--656",
    year = "2003"
}

@article{Schwenk:2002fq,
    author = "Schwenk, Achim and Friman, Bengt and Brown, Gerald E.",
    title = "{Renormalization group approach to neutron matter: Quasiparticle interactions, superfluid gaps and the equation of state}",
    eprint = "nucl-th/0207004",
    archivePrefix = "arXiv",
    doi = "10.1016/S0375-9474(02)01290-3",
    journal = "Nucl. Phys. A",
    volume = "713",
    pages = "191--216",
    year = "2003"
}

@article{Chen:1993bam,
    author = "Chen, J. M. C. and Clark, J. W. and Dav\'e, R. D. and Khodel, V. V.",
    title = "{Pairing gaps in nucleonic superfluids}",
    doi = "10.1016/0375-9474(93)90314-N",
    journal = "Nucl. Phys. A",
    volume = "555",
    pages = "59--89",
    year = "1993"
}

@article{Amundsen:1984hua,
    author = "Amundsen, L. and Ostgaard, E.",
    title = "{Superfluidity of neutron matter (II). triplet pairing}",
    reportNumber = "TRONDHEIM-84-16",
    doi = "10.1016/S0375-9474(85)80012-9",
    journal = "Nucl. Phys. A",
    volume = "442",
    pages = "163--188",
    year = "1985"
}

@article{Akmal:1998cf,
    author = "Akmal, A. and Pandharipande, V. R. and Ravenhall, D. G.",
    title = "{The Equation of state of nucleon matter and neutron star structure}",
    eprint = "nucl-th/9804027",
    archivePrefix = "arXiv",
    doi = "10.1103/PhysRevC.58.1804",
    journal = "Phys. Rev. C",
    volume = "58",
    pages = "1804--1828",
    year = "1998"
}

@article{Wiringa:1994wb,
    author = "Wiringa, Robert B. and Stoks, V. G. J. and Schiavilla, R.",
    title = "{An Accurate nucleon-nucleon potential with charge independence breaking}",
    eprint = "nucl-th/9408016",
    archivePrefix = "arXiv",
    reportNumber = "PHY-7742-TH-94, CEBAF-TH-94-19",
    doi = "10.1103/PhysRevC.51.38",
    journal = "Phys. Rev. C",
    volume = "51",
    pages = "38--51",
    year = "1995"
}

@article{Machleidt:1995km,
    author = "Machleidt, R. and Sammarruca, F. and Song, Y.",
    title = "{Nonlocal nature of the nuclear force and its impact on nuclear structure}",
    eprint = "nucl-th/9510023",
    archivePrefix = "arXiv",
    doi = "10.1103/PhysRevC.53.R1483",
    journal = "Phys. Rev. C",
    volume = "53",
    number = "4",
    pages = "R1483--R1487",
    year = "1996"
}

@article{Wiringa:1988tp,
    author = "Wiringa, Robert B. and Fiks, V. and Fabrocini, A.",
    title = "{Equation of state for dense nucleon matter}",
    doi = "10.1103/PhysRevC.38.1010",
    journal = "Phys. Rev. C",
    volume = "38",
    pages = "1010--1037",
    year = "1988"
}

@article{Akmal:1997ft,
    author = "Akmal, A. and Pandharipande, V. R.",
    title = "{Spin - isospin structure and pion condensation in nucleon matter}",
    eprint = "nucl-th/9705013",
    archivePrefix = "arXiv",
    doi = "10.1103/PhysRevC.56.2261",
    journal = "Phys. Rev. C",
    volume = "56",
    pages = "2261--2279",
    year = "1997"
}

@article{Typel:2009sy,
    author = "Typel, S. and Ropke, G. and Klahn, T. and Blaschke, D. and Wolter, H. H.",
    title = "{Composition and thermodynamics of nuclear matter with light clusters}",
    eprint = "0908.2344",
    archivePrefix = "arXiv",
    primaryClass = "nucl-th",
    doi = "10.1103/PhysRevC.81.015803",
    journal = "Phys. Rev. C",
    volume = "81",
    pages = "015803",
    year = "2010"
}

@article{Tolos:2016hhl,
    author = "Tolos, Laura and Centelles, Mario and Ramos, Angels",
    title = "{Equation of State for Nucleonic and Hyperonic Neutron Stars with Mass and Radius Constraints}",
    eprint = "1610.00919",
    archivePrefix = "arXiv",
    primaryClass = "astro-ph.HE",
    doi = "10.3847/1538-4357/834/1/3",
    journal = "Astrophys. J.",
    volume = "834",
    number = "1",
    pages = "3",
    year = "2017"
}

@article{Chen:2014sca,
    author = "Chen, Wei-Chia and Piekarewicz, J.",
    title = "{Building relativistic mean field models for finite nuclei and neutron stars}",
    eprint = "1408.4159",
    archivePrefix = "arXiv",
    primaryClass = "nucl-th",
    doi = "10.1103/PhysRevC.90.044305",
    journal = "Phys. Rev. C",
    volume = "90",
    number = "4",
    pages = "044305",
    year = "2014"
}

@article{Sagun:2013moa,
    author = "Sagun, V. V. and Ivanytskyi, A. I. and Bugaev, K. A. and Mishustin, I. N.",
    title = "{The statistical multifragmentation model for liquid-gas phase transition with a compressible nuclear liquid}",
    eprint = "1306.2372",
    archivePrefix = "arXiv",
    primaryClass = "nucl-th",
    doi = "10.1016/j.nuclphysa.2013.12.012",
    journal = "Nucl. Phys. A",
    volume = "924",
    pages = "24--46",
    year = "2014"
}

@article{Sagun:2018cpi,
    author = "Sagun, Violetta V. and Lopes, Il{\'\i}dio and Ivanytskyi, Aleksei I.",
    title = "{The induced surface tension contribution for the equation of state of neutron stars}",
    eprint = "1805.04976",
    archivePrefix = "arXiv",
    primaryClass = "astro-ph.HE",
    doi = "10.3847/1538-4357/aaf805",
    journal = "Astrophys. J.",
    volume = "871",
    number = "2",
    pages = "157",
    year = "2019"
}

@article{Cooper:1956zz,
    author = "Cooper, Leon N.",
    title = "{Bound electron pairs in a degenerate Fermi gas}",
    doi = "10.1103/PhysRev.104.1189",
    journal = "Phys. Rev.",
    volume = "104",
    pages = "1189--1190",
    year = "1956"
}

@article{AngelesPerez-Garcia:2022qzs,
    author = "{\'A}ngeles P{\'e}rez-Garc{\'\i}a, M. and Grigorian, H. and Albertus, C. and Barba, D. and Silk, J.",
    title = "{Cooling of Neutron Stars admixed with light dark matter: A case study}",
    eprint = "2202.00702",
    archivePrefix = "arXiv",
    primaryClass = "hep-ph",
    doi = "10.1016/j.physletb.2022.136937",
    journal = "Phys. Lett. B",
    volume = "827",
    pages = "136937",
    year = "2022"
}

@article{Bohr:1958zz,
    author = "Bohr, A. and Mottelson, B. R. and Pines, D.",
    title = "{Possible Analogy between the Excitation Spectra of Nuclei and Those of the Superconducting Metallic State}",
    doi = "10.1103/PhysRev.110.936",
    journal = "Phys. Rev.",
    volume = "110",
    pages = "936--938",
    year = "1958"
}

@article{Sagun:2016nlv,
    author = "Sagun, V. V. and Bugaiev, K. A. and Ivanytskyi, A. I. and Oliinychenko, D. R. and Mishustin, I. N.",
    editor = "Foka, Y. and Brambilla, N. and Kovalenko, V.",
    title = "{Effects of Induced Surface Tension in Nuclear and Hadron Matter}",
    eprint = "1611.07071",
    archivePrefix = "arXiv",
    primaryClass = "nucl-th",
    doi = "10.1051/epjconf/201713709007",
    journal = "EPJ Web Conf.",
    volume = "137",
    pages = "09007",
    year = "2017"
}

@article{Sagun:2017eye,
    author = "Sagun, V. V. and Bugaev, K. A. and Ivanytskyi, A. I. and Yakimenko, I. P. and Nikonov, E. G. and Taranenko, A. V. and Greiner, C. and Blaschke, D. B. and Zinovjev, G. M.",
    title = "{Hadron Resonance Gas Model with Induced Surface Tension}",
    eprint = "1703.00049",
    archivePrefix = "arXiv",
    primaryClass = "hep-ph",
    doi = "10.1140/epja/i2018-12535-1",
    journal = "Eur. Phys. J. A",
    volume = "54",
    number = "6",
    pages = "100",
    year = "2018"
}

@article{Kucherenko:2021oxg,
    author = "Kucherenko, V. S. and Bugaev, K. A. and Sagun, V. and Ivanytskyi, O.",
    title = "{Statistical Multifragmentation Model within the Extended Morphological Thermodynamics Approach}",
    eprint = "2106.09838",
    archivePrefix = "arXiv",
    primaryClass = "nucl-th",
    doi = "10.15407/ujpe67.9.639",
    journal = "Ukr. J. Phys.",
    volume = "67",
    number = "9",
    pages = "639",
    year = "2022"
}

@article{Sagun:2020qvc,
    author = "Sagun, V. and Panotopoulos, G. and Lopes, I.",
    title = "{Asteroseismology: radial oscillations of neutron stars with realistic equation of state}",
    eprint = "2002.12209",
    archivePrefix = "arXiv",
    primaryClass = "astro-ph.HE",
    doi = "10.1103/PhysRevD.101.063025",
    journal = "Phys. Rev. D",
    volume = "101",
    number = "6",
    pages = "063025",
    year = "2020"
}

@article{Horowitz:2001ya,
    author = "Horowitz, Charles J. and Piekarewicz, Jorge",
    title = "{The Neutron radii of Pb-208 and neutron stars}",
    eprint = "nucl-th/0108036",
    archivePrefix = "arXiv",
    doi = "10.1103/PhysRevC.64.062802",
    journal = "Phys. Rev. C",
    volume = "64",
    pages = "062802",
    year = "2001"
}

@article{Lalazissis:1996rd,
    author = "Lalazissis, G. A. and Konig, J. and Ring, P.",
    title = "{A New parametrization for the Lagrangian density of relativistic mean field theory}",
    eprint = "nucl-th/9607039",
    archivePrefix = "arXiv",
    doi = "10.1103/PhysRevC.55.540",
    journal = "Phys. Rev. C",
    volume = "55",
    pages = "540--543",
    year = "1997"
}

@article{Pais:2016xiu,
    author = "Pais, Helena and Provid{\^e}ncia, Constan{\c{c}}a",
    title = "{Vlasov formalism for extended relativistic mean field models: The crust-core transition and the stellar matter equation of state}",
    eprint = "1607.05899",
    archivePrefix = "arXiv",
    primaryClass = "nucl-th",
    doi = "10.1103/PhysRevC.94.015808",
    journal = "Phys. Rev. C",
    volume = "94",
    number = "1",
    pages = "015808",
    year = "2016"
}

@article{Carriere:2002bx,
    author = "Carriere, J. and Horowitz, C. J. and Piekarewicz, J.",
    title = "{Low mass neutron stars and the equation of state of dense matter}",
    eprint = "nucl-th/0211015",
    archivePrefix = "arXiv",
    doi = "10.1086/376515",
    journal = "Astrophys. J.",
    volume = "593",
    pages = "463--471",
    year = "2003"
}

@article{Negele:1971vb,
    author = "Negele, John W. and Vautherin, D.",
    title = "{Neutron star matter at subnuclear densities}",
    doi = "10.1016/0375-9474(73)90349-7",
    journal = "Nucl. Phys. A",
    volume = "207",
    pages = "298--320",
    year = "1973"
}

@article{1990A&A...227..431H,
       author = "Haensel, P. and Zdunik, J.~L.",
        title = "{Non-equilibrium processes in the crust of an accreting neutron star}",
      journal = "Astron. Astrophys",
         year = "1990",
       volume = "227",
       number = "2",
        pages = "431--436",
       adsurl = {https://ui.adsabs.harvard.edu/abs/1990A&A...227..431H}
}

@article{Buschmann:2021juv,
    author = "Buschmann, Malte and Dessert, Christopher and Foster, Joshua W. and Long, Andrew J. and Safdi, Benjamin R.",
    title = "{Upper Limit on the QCD Axion Mass from Isolated Neutron Star Cooling}",
    eprint = "2111.09892",
    archivePrefix = "arXiv",
    primaryClass = "hep-ph",
    doi = "10.1103/PhysRevLett.128.091102",
    journal = "Phys. Rev. Lett.",
    volume = "128",
    number = "9",
    pages = "091102",
    year = "2022"
}

@article{PValuesNuisance,
author = {Demortier, Luc},
year = {2007},
month = {01},
pages = {},
title = {P Values and Nuisance Parameters},
journal = {PHYSTAT LHC Workshop on Statistical Issues for LHC Physics, PHYSTAT 2007 - Proceedings}
}

@article{Ho:2009mm,
    author = "Ho, Wynn C. G. and Heinke, Craig O.",
    title = "{A Neutron Star with a Carbon Atmosphere in the Cassiopeia A Supernova Remnant}",
    eprint = "0911.0672",
    archivePrefix = "arXiv",
    primaryClass = "astro-ph.HE",
    doi = "10.1038/nature08525",
    journal = "Nature",
    volume = "462",
    pages = "71--73",
    year = "2009"
}

@article{Heinke:2010cr,
    author = "Heinke, Craig O. and Ho, Wynn C. G.",
    title = "{Direct Observation of the Cooling of the Cassiopeia A Neutron Star}",
    eprint = "1007.4719",
    archivePrefix = "arXiv",
    primaryClass = "astro-ph.HE",
    doi = "10.1088/2041-8205/719/2/L167",
    journal = "Astrophys. J. Lett.",
    volume = "719",
    pages = "L167--L171",
    year = "2010"
}

@article{Shternin:2010qi,
    author = "Shternin, Peter S. and Yakovlev, Dmitry G. and Heinke, Craig O. and Ho, Wynn C. G. and Patnaude, Daniel J.",
    title = "{Cooling neutron star in the Cassiopeia\textasciitilde{}A supernova remnant: Evidence for superfluidity in the core}",
    eprint = "1012.0045",
    archivePrefix = "arXiv",
    primaryClass = "astro-ph.SR",
    doi = "10.1111/j.1745-3933.2011.01015.x",
    journal = "Mon. Not. Roy. Astron. Soc.",
    volume = "412",
    pages = "L108--L112",
    year = "2011"
}

@article{Elshamouty:2013nfa,
    author = "Elshamouty, K. G. and Heinke, C. O. and Sivakoff, G. R. and Ho, W. C. G. and Shternin, P. S. and Yakovlev, D. G. and Patnaude, D. J. and David, L.",
    title = "{Measuring the Cooling of the Neutron Star in Cassiopeia A with all Chandra X-ray Observatory Detectors}",
    eprint = "1306.3387",
    archivePrefix = "arXiv",
    primaryClass = "astro-ph.HE",
    doi = "10.1088/0004-637X/777/1/22",
    journal = "Astrophys. J.",
    volume = "777",
    pages = "22",
    year = "2013"
}

@article{Wijngaarden:2019tht,
    author = "Wijngaarden, M. J. P. and Ho, Wynn C. G. and Chang, Philip and Heinke, Craig O. and Page, Dany and Beznogov, Mikhail and Patnaude, Daniel J.",
    title = "{Diffusive nuclear burning in cooling simulations and application to new temperature data of the Cassiopeia A neutron star}",
    eprint = "1901.01012",
    archivePrefix = "arXiv",
    primaryClass = "astro-ph.HE",
    doi = "10.1093/mnras/stz042",
    journal = "Mon. Not. Roy. Astron. Soc.",
    volume = "484",
    number = "1",
    pages = "974--988",
    year = "2019"
}

@article{Shternin:2022rti,
    author = "Shternin, Peter S. and Ofengeim, Dmitry D. and Heinke, Craig O. and Ho, Wynn C. G.",
    title = "{Constraints on neutron star superfluidity from the cooling neutron star in Cassiopeia~A using all Chandra ACIS-S observations}",
    eprint = "2211.02526",
    archivePrefix = "arXiv",
    primaryClass = "astro-ph.HE",
    doi = "10.1093/mnras/stac3226",
    journal = "Mon. Not. Roy. Astron. Soc.",
    volume = "518",
    number = "2",
    pages = "2775--2793",
    year = "2022"
}

@article{Page:2010aw,
    author = "Page, Dany and Prakash, Madappa and Lattimer, James M. and Steiner, Andrew W.",
    title = "{Rapid Cooling of the Neutron Star in Cassiopeia A Triggered by Neutron Superfluidity in Dense Matter}",
    eprint = "1011.6142",
    archivePrefix = "arXiv",
    primaryClass = "astro-ph.HE",
    doi = "10.1103/PhysRevLett.106.081101",
    journal = "Phys. Rev. Lett.",
    volume = "106",
    pages = "081101",
    year = "2011"
}

@article{Beznogov:2015ewa,
    author = "Beznogov, M. V. and Yakovlev, D. G.",
    title = "{Statistical theory of thermal evolution of neutron stars {\textendash} II. Limitations on direct Urca threshold}",
    eprint = "1507.04206",
    archivePrefix = "arXiv",
    primaryClass = "astro-ph.SR",
    doi = "10.1093/mnras/stv1293",
    journal = "Mon. Not. Roy. Astron. Soc.",
    volume = "452",
    number = "1",
    pages = "540--548",
    year = "2015"
}

@article{Wei:2018yaj,
    author = "Wei, J. -B. and Burgio, G. F. and Schulze, H. -J.",
    title = "{Neutron star cooling with microscopic equations of state}",
    eprint = "1812.07306",
    archivePrefix = "arXiv",
    primaryClass = "astro-ph.HE",
    doi = "10.1093/mnras/stz336",
    journal = "Mon. Not. Roy. Astron. Soc.",
    volume = "484",
    pages = "5162",
    year = "2019"
}

@article{Wei:2020xdk,
    author = "Wei, Jin-Biao and Burgio, Fiorella and Schulze, Hans-Josef",
    title = "{Nuclear Pairing Gaps and Neutron Star Cooling}",
    eprint = "2010.03916",
    archivePrefix = "arXiv",
    primaryClass = "nucl-th",
    doi = "10.3390/universe6080115",
    journal = "Universe",
    volume = "6",
    number = "8",
    pages = "115",
    year = "2020"
}

@article{Dietrich:2020efo,
    author = "Dietrich, Tim and Coughlin, Michael W. and Pang, Peter T. H. and Bulla, Mattia and Heinzel, Jack and Issa, Lina and Tews, Ingo and Antier, Sarah",
    title = "{Multimessenger constraints on the neutron-star equation of state and the Hubble constant}",
    eprint = "2002.11355",
    archivePrefix = "arXiv",
    primaryClass = "astro-ph.HE",
    reportNumber = "LA-UR-20-21470",
    doi = "10.1126/science.abb4317",
    journal = "Science",
    volume = "370",
    number = "6523",
    pages = "1450--1453",
    year = "2020"
}

@article{Tsiopelas:2020nzm,
    author = "Tsiopelas, Stefanos and Sagun, Violetta",
    title = "{Neutron star cooling within the equation of state with induced surface tension}",
    eprint = "2006.06351",
    archivePrefix = "arXiv",
    primaryClass = "astro-ph.HE",
    doi = "10.3390/particles3040045",
    journal = "Particles",
    volume = "3",
    number = "4",
    pages = "693--705",
    year = "2020"
}

@article{Tsiopelas:2021ayd,
    author = "Tsiopelas, Stefanos and Sagun, Violetta",
    editor = "Zen Vasconcellos, C. A. and Hess, P. O. and Piccinelli, G. and Vargas Maga{\~n}a, M. and Ure{\~n}a-Lopez, L. and Gonz{\'a}lez Felipe, R. and Boller, Th. and Gullberg, S.",
    title = "{Thermal evolution of neutron stars described within the equation of state with induced surface tension}",
    doi = "10.1002/asna.202113929",
    journal = "Astron. Nachr.",
    volume = "342",
    number = "1-2",
    pages = "332--336",
    year = "2021"
}

@article{Sedrakian:2018ydt,
    author = "Sedrakian, Armen and Clark, John W.",
    title = "{Superfluidity in nuclear systems and neutron stars}",
    eprint = "1802.00017",
    archivePrefix = "arXiv",
    primaryClass = "nucl-th",
    doi = "10.1140/epja/i2019-12863-6",
    journal = "Eur. Phys. J. A",
    volume = "55",
    number = "9",
    pages = "167",
    year = "2019"
}

@article{Beznogov:2021ijc,
    author = "Beznogov, M. V. and Potekhin, A. Y. and Yakovlev, D. G.",
    title = "{Heat blanketing envelopes of neutron stars}",
    eprint = "2103.12422",
    archivePrefix = "arXiv",
    primaryClass = "astro-ph.SR",
    doi = "10.1016/j.physrep.2021.03.004",
    journal = "Phys. Rept.",
    volume = "919",
    pages = "1--68",
    year = "2021"
}

@article{Blaschke:2006gd,
    author = "Blaschke, David and Grigorian, H.",
    editor = "Faessler, Amand",
    title = "{Unmasking neutron star interiors using cooling simulations}",
    eprint = "astro-ph/0612092",
    archivePrefix = "arXiv",
    doi = "10.1016/j.ppnp.2006.12.035",
    journal = "Prog. Part. Nucl. Phys.",
    volume = "59",
    pages = "139--146",
    year = "2007"
}

@article{Blaschke:2004vq,
    author = "Blaschke, David and Grigorian, H. and Voskresensky, D. N.",
    title = "{Cooling of neutron stars: Hadronic model}",
    eprint = "astro-ph/0403170",
    archivePrefix = "arXiv",
    reportNumber = "MPG-VT-UR-246-04",
    doi = "10.1051/0004-6361:20040404",
    journal = "Astron. Astrophys.",
    volume = "424",
    pages = "979--992",
    year = "2004"
}

@article{Blaschke:2013vma,
    author = "Blaschke, D. and Grigorian, H. and Voskresensky, D. N.",
    title = "{Nuclear medium cooling scenario in the light of new Cas A cooling data and the 2$M_\odot$ pulsar mass measurements}",
    eprint = "1308.4093",
    archivePrefix = "arXiv",
    primaryClass = "nucl-th",
    doi = "10.1103/PhysRevC.88.065805",
    journal = "Phys. Rev. C",
    volume = "88",
    number = "6",
    pages = "065805",
    year = "2013"
}

@article{Baldo:1992kzz,
    author = "Baldo, M. and Cugnon, J. and Lejeune, A. and Lombardo, U.",
    title = "{Proton and neutron superfluidity in neutron star matter}",
    doi = "10.1016/0375-9474(92)90387-Y",
    journal = "Nucl. Phys. A",
    volume = "536",
    pages = "349--365",
    year = "1992"
}

@article{Kunert:2026hmx,
    author = "Kunert, Nina and Grams, Guilherme and Newton, William and Giangrandi, Edoardo and Puecher, Anna and Koehn, Hauke and Sagun, Violetta and Dietrich, Tim",
    title = "{Unified study of nuclear physics and dark matter constraints through gravitational-wave observations of binary neutron star mergers}",
    eprint = "2602.19627",
    archivePrefix = "arXiv",
    primaryClass = "astro-ph.HE",
    doi = "10.1103/9vds-3nmb",
    journal = "Phys. Rev. D",
    volume = "114",
    number = "6",
    pages = "064056",
    year = "2026"
}

@article{Baldo:2011gz,
    author = "Baldo, M. and Burgio, G. F.",
    title = "{Properties of the nuclear medium}",
    eprint = "1102.1364",
    archivePrefix = "arXiv",
    primaryClass = "nucl-th",
    reportNumber = "INFN-CT-02-2011",
    doi = "10.1088/0034-4885/75/2/026301",
    journal = "Rept. Prog. Phys.",
    volume = "75",
    pages = "026301",
    year = "2012"
}

@article{Popov:2004ey,
    author = "Popov, S. and Grigorian, H. and Turolla, R. and Blaschke, David",
    title = "{Population synthesis as a probe of neutron star thermal evolution}",
    eprint = "astro-ph/0411618",
    archivePrefix = "arXiv",
    doi = "10.1051/0004-6361:20042412",
    journal = "Astron. Astrophys.",
    volume = "448",
    pages = "327",
    year = "2006"
}

@article{Taranto:2015ubs,
    author = "Taranto, G. and Burgio, G. F. and Schulze, H. -J.",
    title = "{Cassiopeia A and direct URCA cooling}",
    eprint = "1511.04243",
    archivePrefix = "arXiv",
    primaryClass = "nucl-th",
    doi = "10.1093/mnras/stv2756",
    journal = "Mon. Not. Roy. Astron. Soc.",
    volume = "456",
    number = "2",
    pages = "1451--1458",
    year = "2016"
}

@article{Blaschke:2011gc,
    author = "Blaschke, D. and Grigorian, H. and Voskresensky, D. N. and Weber, F.",
    title = "{On the Cooling of the Neutron Star in Cassiopeia A}",
    eprint = "1108.4125",
    archivePrefix = "arXiv",
    primaryClass = "nucl-th",
    doi = "10.1103/PhysRevC.85.022802",
    journal = "Phys. Rev. C",
    volume = "85",
    pages = "022802",
    year = "2012"
}

@article{Hamaguchi:2018oqw,
    author = "Hamaguchi, Koichi and Nagata, Natsumi and Yanagi, Keisuke and Zheng, Jiaming",
    title = "{Limit on the Axion Decay Constant from the Cooling Neutron Star in Cassiopeia A}",
    eprint = "1806.07151",
    archivePrefix = "arXiv",
    primaryClass = "hep-ph",
    reportNumber = "UT-18-13, IPMU 18-0111, IPMU-18-0111",
    doi = "10.1103/PhysRevD.98.103015",
    journal = "Phys. Rev. D",
    volume = "98",
    number = "10",
    pages = "103015",
    year = "2018"
}

@misc{Neutronstarcoolingcode,
author = {{Page}, Dany},
title = "{NSCool: Neutron star cooling code}",
howpublished = {Astrophysics Source Code Library, record ascl:1609.009},
year = 2016,
eid = {ascl:1609.009},
pages = {ascl:1609.009},
archivePrefix = {ascl},
eprint = {1609.009},
adsurl = {https://ui.adsabs.harvard.edu/abs/2013ascl.soft05002P}
}

@article{Sedrakian:2015krq,
    author = "Sedrakian, Armen",
    title = "{Axion cooling of neutron stars}",
    eprint = "1512.07828",
    archivePrefix = "arXiv",
    primaryClass = "astro-ph.HE",
    doi = "10.1103/PhysRevD.93.065044",
    journal = "Phys. Rev. D",
    volume = "93",
    number = "6",
    pages = "065044",
    year = "2016"
}

@article{Sedrakian:2018kdm,
    author = "Sedrakian, Armen",
    title = "{Axion cooling of neutron stars. II. Beyond hadronic axions}",
    eprint = "1810.00190",
    archivePrefix = "arXiv",
    primaryClass = "astro-ph.HE",
    doi = "10.1103/PhysRevD.99.043011",
    journal = "Phys. Rev. D",
    volume = "99",
    number = "4",
    pages = "043011",
    year = "2019"
}

@article{Becker:2012dj,
    author = "Becker, Werner and Prinz, Tobias and Winkler, P. Frank and Petre, Robert",
    title = "{The Proper Motion of the Central Compact Object RX J0822-4300 in the Supernova Remnant Puppis A}",
    eprint = "1204.3510",
    archivePrefix = "arXiv",
    primaryClass = "astro-ph.HE",
    doi = "10.1088/0004-637X/755/2/141",
    journal = "Astrophys. J.",
    volume = "755",
    pages = "141",
    year = "2012"
}

@article{DeLuca:2011yr,
    author = "De Luca, A. and others",
    title = "{A time-variable, phase-dependent emission line in the X-ray spectrum of the isolated neutron star RX J0822-4300}",
    eprint = "1112.4602",
    archivePrefix = "arXiv",
    primaryClass = "astro-ph.HE",
    doi = "10.1111/j.1745-3933.2011.01209.x",
    journal = "Mon. Not. Roy. Astron. Soc.",
    volume = "421",
    pages = "72",
    year = "2012"
}

@article{Allen:2014yra,
    author = "Allen, G. E. and Chow, K. and DeLaney, T. and Filipovic, M. D. and Houck, J. C. and Pannuti, T. G. and Stage, M. D.",
    title = "{On the Expansion Rate, Age, and Distance of the Supernova Remnant G266.2-1.2 (Vela Jr.)}",
    eprint = "1410.7435",
    archivePrefix = "arXiv",
    primaryClass = "astro-ph.HE",
    doi = "10.1088/0004-637X/798/2/82",
    journal = "Astrophys. J.",
    volume = "798",
    number = "2",
    pages = "82",
    year = "2015"
}

@article{Ho:2026guu,
    author = "Ho, Wynn C. G. and Simkhayeva, Esther and Potekhin, Alexander Y.",
    title = "{Neutron star cooling implications and magnetic field of the Vela Junior central compact object from all XMM{\textendash}Newton and Chandra spectra}",
    eprint = "2601.14388",
    archivePrefix = "arXiv",
    primaryClass = "astro-ph.HE",
    doi = "10.1093/mnras/stag126",
    journal = "Mon. Not. Roy. Astron. Soc.",
    volume = "546",
    number = "3",
    pages = "stag126",
    year = "2026"
}

@article{Pires:2015gna,
    author = "Pires, Adriana M. and Motch, Christian and Turolla, Roberto and Popov, Sergei B. and Schwope, Axel D. and Treves, Aldo",
    title = "{New XMM-Newton observation of the thermally emitting isolated neutron star 2XMM J104608.7-594306}",
    eprint = "1508.05246",
    archivePrefix = "arXiv",
    primaryClass = "astro-ph.HE",
    doi = "10.1051/0004-6361/201526436",
    journal = "Astron. Astrophys.",
    volume = "583",
    pages = "A117",
    year = "2015"
}

@article{Mereghetti:2002za,
    author = "Mereghetti, S. and De Luca, Andrew and Caraveo, P. A. and Becker, W. and Mignani, R. and Bignami, G. F.",
    title = "{Pulse phase variations of the x-ray spectral features in the radio-quiet neutron star 1e 1207-5209}",
    eprint = "astro-ph/0207296",
    archivePrefix = "arXiv",
    doi = "10.1086/344435",
    journal = "Astrophys. J.",
    volume = "581",
    pages = "1280--1285",
    year = "2002"
}

@article{Roger:1988,
author = {Roger, R. and Milne, D. and Kesteven, Michael and Wellington, K. and Haynes, Roslynn},
year = {1988},
month = {08},
pages = {940},
title = {Symmetry of the radio emission from two high-latitude supernova remnants, G296.5 + 10.0 and G327.6 + 14.6 (SN 1006)},
volume = {332},
journal = {Astrophys. J.},
doi = {10.1086/166703}
}

@article{Borkowski:2018omn,
    author = "Borkowski, Kazimierz J. and Reynolds, Stephen P. and Williams, Brian J. and Petre, Robert",
    title = "{Expansion and Age of the X-ray Synchrotron-Dominated Supernova Remnant G330.2+1.0}",
    eprint = "1811.01998",
    archivePrefix = "arXiv",
    primaryClass = "astro-ph.HE",
    doi = "10.3847/2041-8213/aaedb5",
    journal = "Astrophys. J. Lett.",
    volume = "868",
    number = "2",
    pages = "L21",
    year = "2018"
}

@article{Ho:2021hwy,
    author = "Ho, Wynn C. G. and Zhao, Yue and Heinke, Craig O. and Kaplan, D. L. and Shternin, Peter S. and Wijngaarden, M. J. P.",
    title = "{X-ray bounds on cooling, composition, and magnetic field of the Cassiopeia~A neutron star and young central compact objects}",
    eprint = "2107.08060",
    archivePrefix = "arXiv",
    primaryClass = "astro-ph.HE",
    doi = "10.1093/mnras/stab2081",
    journal = "Mon. Not. Roy. Astron. Soc.",
    volume = "506",
    number = "4",
    pages = "5015--5029",
    year = "2021"
}

@article{Cassam-Chenai:2004jon,
    author = "Cassam-Chenai, Gamil and Decourchelle, A. and Ballet, J. and Sauvageot, J. -L. and Dubner, G. and Giacani, E.",
    title = "{XMM-Newton observations of the supernova remnant RX J1713.7-3946 and its central source}",
    eprint = "astro-ph/0407333",
    archivePrefix = "arXiv",
    doi = "10.1051/0004-6361:20041154",
    journal = "Astron. Astrophys.",
    volume = "427",
    pages = "199",
    year = "2004"
}

@article{Mayer:2021wvs,
    author = "Mayer, Martin G. F. and Becker, Werner",
    title = "{A kinematic study of central compact objects and their host supernova remnants}",
    eprint = "2106.00700",
    archivePrefix = "arXiv",
    primaryClass = "astro-ph.HE",
    doi = "10.1051/0004-6361/202141119",
    journal = "Astron. Astrophys.",
    volume = "651",
    pages = "A40",
    year = "2021"
}

@article{Maxted:2017xfs,
    author = {Maxted, Nigel and Burton, Michael and Braiding, Catherine and Rowell, Gavin and Sano, Hidetoshi and Voisin, Fabien and Capasso, Massimo and P{\"u}hlhofer, Gerd and Fukui, Yasuo},
    title = "{Probing the local environment of the supernova remnant HESS J1731{\ensuremath{-}}347 with CO and CS observations}",
    eprint = "1710.06101",
    archivePrefix = "arXiv",
    primaryClass = "astro-ph.HE",
    doi = "10.1093/mnras/stx2727",
    journal = "Mon. Not. Roy. Astron. Soc.",
    volume = "474",
    number = "1",
    pages = "662--676",
    year = "2018"
}

@article{Doroshenko:2022nwp,
    author = {Doroshenko, Victor and Suleimanov, Valery and P{\"u}hlhofer, Gerd and Santangelo, Andrea},
    title = "{A strangely light neutron star within a supernova remnant}",
    doi = "10.1038/s41550-022-01800-1",
    journal = "Nature Astron.",
    volume = "6",
    number = "12",
    pages = "1444--1451",
    year = "2022"
}

@article{Sasaki:2018wnz,
    author = {Sasaki, Manami and M{\"a}kel{\"a}, Minja M. and Klochkov, Dmitry and Santangelo, Andrea and Suleimanov, Valery},
    title = "{Infrared and X-ray study of the Galactic SNR G15.9+0.2}",
    eprint = "1806.11365",
    archivePrefix = "arXiv",
    primaryClass = "astro-ph.HE",
    doi = "10.1093/mnras/sty1596",
    journal = "Mon. Not. Roy. Astron. Soc.",
    volume = "479",
    number = "3",
    pages = "3033--3041",
    year = "2018"
}

@article{Klochkov:2016lly,
    author = "Klochkov, D. and Suleimanov, V. and Sasaki, M. and Santangelo, A.",
    title = "{Study of a new central compact object: The neutron star in the supernova remnant G15.9+0.2}",
    eprint = "1607.08021",
    archivePrefix = "arXiv",
    primaryClass = "astro-ph.HE",
    doi = "10.1051/0004-6361/201629208",
    journal = "Astron. Astrophys.",
    volume = "592",
    pages = "L12",
    year = "2016"
}

@article{Sun:2004hz,
    author = "Sun, Ming and Seward, F. D. and Smith, R. K. and Slane, P. O.",
    title = "{Chandra view of Kes 79: A Nearly isothermal SNR with rich spatial structure}",
    eprint = "astro-ph/0401165",
    archivePrefix = "arXiv",
    doi = "10.1086/382666",
    journal = "Astrophys. J.",
    volume = "605",
    pages = "742--750",
    year = "2004"
}

@article{Bogdanov:2014hna,
    author = "Bogdanov, Slavko",
    title = "{Modeling the X-rays from the Central Compact Object PSR J1852+0040 in Kesteven 79: Evidence for a Strongly Magnetized Neutron Star}",
    eprint = "1406.0515",
    archivePrefix = "arXiv",
    primaryClass = "astro-ph.HE",
    doi = "10.1088/0004-637X/790/2/94",
    journal = "Astrophys. J.",
    volume = "790",
    number = "2",
    pages = "94",
    year = "2014"
}

@article{Kirichenko:2014ona,
    author = "Kirichenko, Aida and Danilenko, Andrey and Shibanov, Yury and Shternin, Peter and Zharikov, Sergey and Zyuzin, Dmitry",
    title = "{Deep optical observations of the $\gamma$-ray pulsar J0357+3205}",
    eprint = "1402.2246",
    archivePrefix = "arXiv",
    primaryClass = "astro-ph.SR",
    doi = "10.1051/0004-6361/201322628",
    journal = "Astron. Astrophys.",
    volume = "564",
    pages = "A81",
    year = "2014"
}

@article{Ng:2006vh,
    author = "Ng, Chi-Y. and Romani, Roger W. and Brisken, Walter F. and Chatterjee, Shami and Kramer, Michael",
    title = "{The Origin and Motion of PSR J0538+2817 in S147}",
    eprint = "astro-ph/0611068",
    archivePrefix = "arXiv",
    doi = "10.1086/510576",
    journal = "Astrophys. J.",
    volume = "654",
    pages = "487--493",
    year = "2006"
}

@article{Kramer:2003au,
    author = "Kramer, Michael and Lyne, A. G. and Hobbs, G. and Lohmer, O. and Carr, P. and Jordan, C. and Wolszczan, A.",
    title = "{Proper motion, age and initial spin period of PSR J0538+2817 in S147}",
    eprint = "astro-ph/0306628",
    archivePrefix = "arXiv",
    doi = "10.1086/378082",
    journal = "Astrophys. J. Lett.",
    volume = "593",
    pages = "L31--L34",
    year = "2003"
}

@article{Chevalier:1998ew,
    author = "Chevalier, Roger A.",
    title = "{Supernova remnants in molecular clouds}",
    eprint = "astro-ph/9805315",
    archivePrefix = "arXiv",
    doi = "10.1086/306710",
    journal = "Astrophys. J.",
    volume = "511",
    pages = "798",
    year = "1999"
}

@article{Swartz:2015fwa,
    author = "Swartz, Douglas A. and Pavlov, George G. and Clarke, Tracy and Castelletti, Gabriela and Zavlin, Vyacheslav E. and Bucciantini, Niccol{\`o} and Karovska, Margarita and van der Horst, Alexander J. and Yukita, Mihoko and Weisskopf, Martin C.",
    title = "{High Spatial Resolution X-ray Spectroscopy of the ic 443 Pulsar Wind Nebula and Environs}",
    eprint = "1506.05507",
    archivePrefix = "arXiv",
    primaryClass = "astro-ph.HE",
    doi = "10.1088/0004-637X/808/1/84",
    journal = "Astrophys. J.",
    volume = "808",
    number = "1",
    pages = "84",
    year = "2015"
}

@inproceedings{Aschenbach:2002es,
    author = "Aschenbach, B.",
    title = "{X-rays from supernova remnants}",
    booktitle = "{270th WE-Heraeus Seminar on Neutron Stars, Pulsars and Supernova Remnants}",
    eprint = "astro-ph/0208492",
    archivePrefix = "arXiv",
    reportNumber = "MPE-REPORT-278",
    month = "8",
    year = "2002"
}

@article{Ofengeim:2018qkr,
    author = "Ofengeim, Dmitry D. and Zyuzin, Dmitry A.",
    editor = "Sedrakian, Armen",
    title = "{Thermal Spectrum and Neutrino Cooling Rate of the Vela Pulsar}",
    doi = "10.3390/particles1010014",
    journal = "Particles",
    volume = "1",
    number = "1",
    pages = "194--202",
    year = "2018"
}

@article{Igoshev:2018fcp,
    author = "Igoshev, Andrei P.",
    title = "{Ages of radio pulsar: long-term magnetic field evolution}",
    eprint = "1810.12922",
    archivePrefix = "arXiv",
    primaryClass = "astro-ph.HE",
    doi = "10.1093/mnras/sty2945",
    journal = "Mon. Not. Roy. Astron. Soc.",
    volume = "482",
    number = "3",
    pages = "3415--3425",
    year = "2019"
}

@article{Abramkin:2021fzy,
    author = "Abramkin, Vadim and Pavlov, George G. and Shibanov, Yuriy and Kargaltsev, Oleg",
    title = "{Thermal and Nonthermal Emission in the Optical-UV Spectrum of PSR B0950+08*}",
    eprint = "2111.08801",
    archivePrefix = "arXiv",
    primaryClass = "astro-ph.HE",
    doi = "10.3847/1538-4357/ac3a6f",
    journal = "Astrophys. J.",
    volume = "924",
    number = "2",
    pages = "128",
    year = "2022"
}

@article{Migliazzo:2002vd,
    author = "Migliazzo, J. M. and Gaensler, B. M. and Backer, D. C. and Stappers, B. W. and van der Swaluw, E. and Strom, R. G.",
    title = "{Proper motion measurements of pulsar B1951+32 in the supernova remnant CTB 80}",
    eprint = "astro-ph/0202063",
    archivePrefix = "arXiv",
    doi = "10.1086/340002",
    journal = "Astrophys. J. Lett.",
    volume = "567",
    pages = "L141",
    year = "2002"
}

@article{Li:2005qa,
    author = "Li, Xiang-Hua and Lu, F. J. and Li, T. P.",
    title = "{X-ray spectroscopy of PSR B1951+32 and its pulsar wind nebula}",
    eprint = "astro-ph/0504293",
    archivePrefix = "arXiv",
    doi = "10.1086/430941",
    journal = "Astrophys. J.",
    volume = "628",
    pages = "931--937",
    year = "2005"
}

@article{Kumar:2012dm,
    author = "Kumar, Harsha S. and Safi-Harb, Samar and Gonzalez, Marjorie",
    title = "{Chandra and XMM-Newton studies of the supernova remnant G292.2-0.5 associated with the pulsar J1119-6127}",
    eprint = "1205.5766",
    archivePrefix = "arXiv",
    primaryClass = "astro-ph.HE",
    doi = "10.1088/0004-637X/754/2/96",
    journal = "Astrophys. J.",
    volume = "754",
    pages = "96",
    year = "2012"
}

@article{Ng:2012dv,
    author = "Ng, C. -Y. and Kaspi, V. M. and Ho, W. C. G. and Weltevrede, P. and Bogdanov, S. and Shannon, R. and Gonzalez, M. E.",
    title = "{Deep X-ray Observations of the Young High-Magnetic-Field Radio Pulsar J1119-6127 and Supernova Remnant G292.2-0.5}",
    eprint = "1211.2761",
    archivePrefix = "arXiv",
    primaryClass = "astro-ph.HE",
    doi = "10.1088/0004-637X/761/1/65",
    journal = "Astrophys. J.",
    volume = "761",
    pages = "65",
    year = "2012"
}

@article{Tetzlaff:2011kh,
    author = "Tetzlaff, Nina and Eisenbeiss, Thomas and Neuhaeuser, Ralph and Hohle, Markus Matthias",
    title = "{The origin of RXJ1856.5-3754 and RXJ0720.4-3125 -- updated using new parallax measurements}",
    eprint = "1107.1673",
    archivePrefix = "arXiv",
    primaryClass = "astro-ph.GA",
    doi = "10.1111/j.1365-2966.2011.19302.x",
    journal = "Mon. Not. Roy. Astron. Soc.",
    volume = "417",
    pages = "617",
    year = "2011"
}

@article{Hambaryan:2017wvm,
    author = {Hambaryan, V. and Suleimanov, V. and Haberl, F. and Schwope, A. D. and Neuh{\"a}user, R. and Hohle, M. and Werner, K.},
    title = "{The compactness of the isolated neutron star RX J0720.4{\ensuremath{-}}3125}",
    eprint = "1702.07635",
    archivePrefix = "arXiv",
    primaryClass = "astro-ph.HE",
    doi = "10.1051/0004-6361/201630368",
    journal = "Astron. Astrophys.",
    volume = "601",
    pages = "A108",
    year = "2017"
}

@article{Tetzlaff:2012rz,
    author = "Tetzlaff, Nina and Schmidt, Janos G. and Hohle, Markus M. and Neuhaeuser, Ralph",
    title = "{Neutron stars from young nearby associations the origin of RXJ1605.3+3249}",
    eprint = "1202.1388",
    archivePrefix = "arXiv",
    primaryClass = "astro-ph.GA",
    doi = "10.1071/AS11057",
    journal = "Publ. Astron. Soc. Austral.",
    volume = "29",
    pages = "98",
    year = "2012"
}

@article{Malacaria:2019zqr,
    author = "Malacaria, Christian and others",
    title = "{A joint NICER and XMM-Newton view of the ''Magnificent'' thermally emitting X-ray Isolated Neutron Star RX J1605.3+3249}",
    eprint = "1906.02806",
    archivePrefix = "arXiv",
    primaryClass = "astro-ph.HE",
    doi = "10.3847/1538-4357/ab2875",
    journal = "Astrophys. J.",
    volume = "880",
    number = "2",
    pages = "74",
    year = "2019"
}

@article{Pires:2019qsk,
    author = "Pires, A. M. and Schwope, A. D. and Haberl, F. and Zavlin, V. E. and Motch, C. and Zane, S.",
    title = "{A deep XMM-Newton look on the thermally emitting isolated neutron star RX J1605.3+3249}",
    eprint = "1901.08533",
    archivePrefix = "arXiv",
    primaryClass = "astro-ph.HE",
    doi = "10.1051/0004-6361/201834801",
    journal = "Astron. Astrophys.",
    volume = "623",
    pages = "A73",
    year = "2019"
}

@article{Mignani:2012mm,
    author = "Mignani, R. P. and Putte, D. Vande and Cropper, M. and Turolla, R. and Zane, S. and Pellizza, L. J. and Bignone, L. A. and Sartore, N. and Treves, A.",
    title = "{The birthplace and age of the isolated neutron star RX J1856.5-3754}",
    eprint = "1212.3141",
    archivePrefix = "arXiv",
    primaryClass = "astro-ph.HE",
    doi = "10.1093/mnras/sts627",
    journal = "Mon. Not. Roy. Astron. Soc.",
    volume = "429",
    pages = "3517",
    year = "2013"
}

@article{Yoneyama:2017xth,
    author = "Yoneyama, Tomokage and Hayashida, Kiyoshi and Nakajima, Hiroshi and Inoue, Shota and Tsunemi, Hiroshi",
    title = "{Discovery of a keV-X-ray excess in RX J1856.5{\textendash}3754}",
    eprint = "1703.05995",
    archivePrefix = "arXiv",
    primaryClass = "astro-ph.HE",
    doi = "10.1093/pasj/psx025",
    journal = "Publ. Astron. Soc. Jap.",
    volume = "69",
    number = "3",
    pages = "50",
    year = "2017"
}

@article{Kothes:2013hka,
    author = "Kothes, Roland",
    title = "{A Thorough Investigation of Distance and Age of the Pulsar Wind Nebula 3C58}",
    eprint = "1307.8384",
    archivePrefix = "arXiv",
    primaryClass = "astro-ph.GA",
    doi = "10.1051/0004-6361/201219839",
    journal = "Astron. Astrophys.",
    volume = "560",
    pages = "A18",
    year = "2013"
}

@article{Potekhin:2020ttj,
    author = "Potekhin, A. Y. and Zyuzin, D. A. and Yakovlev, D. G. and Beznogov, M. V. and Shibanov, Yu. A.",
    title = "{Thermal luminosities of cooling neutron stars}",
    eprint = "2006.15004",
    archivePrefix = "arXiv",
    primaryClass = "astro-ph.HE",
    doi = "10.1093/mnras/staa1871",
    journal = "Mon. Not. Roy. Astron. Soc.",
    volume = "496",
    number = "4",
    pages = "5052--5071",
    year = "2020"
}

@article{Yakovlev:1999sk,
    author = "Yakovlev, D. G. and Levenfish, K. P. and Shibanov, Yu. A.",
    title = "{Cooling neutron stars and superfluidity in their interiors}",
    eprint = "astro-ph/9906456",
    archivePrefix = "arXiv",
    doi = "10.1070/PU1999v042n08ABEH000556",
    journal = "Phys. Usp.",
    volume = "42",
    pages = "737--778",
    year = "1999"
}

@article{Bardeen:1957mv,
    author = "Bardeen, John and Cooper, L. N. and Schrieffer, J. R.",
    title = "{Theory of superconductivity}",
    doi = "10.1103/PhysRev.108.1175",
    journal = "Phys. Rev.",
    volume = "108",
    pages = "1175--1204",
    year = "1957"
}

@article{Baym:1969fiw,
    author = "Baym, Gordon and Pethick, Christopher and Pines, David",
    title = "{Superfluidity in Neutron Stars}",
    doi = "10.1038/224673a0",
    journal = "Nature",
    volume = "224",
    number = "5220",
    pages = "673--674",
    year = "1969"
}

@article{Chen:1986xzn,
    author = "Chen, J. M. C. and Clark, J. W. and Krotscheck, E. and Smith, R. A.",
    title = "{Nucleonic superfluidity in neutron stars: 1 S 0 neutron pairing in the inner crust}",
    doi = "10.1016/0375-9474(86)90074-6",
    journal = "Nucl. Phys. A",
    volume = "451",
    pages = "509--540",
    year = "1986"
}

@article{Sales:2020aad,
    author = "Sales, Thiago and Louren{\c{c}}o, Odilon and Dutra, Mariana and Negreiros, Rodrigo",
    title = "{Revisiting the thermal relaxation of neutron stars}",
    eprint = "2004.05019",
    archivePrefix = "arXiv",
    primaryClass = "astro-ph.HE",
    doi = "10.1051/0004-6361/202038193",
    journal = "Astron. Astrophys.",
    volume = "642",
    pages = "A42",
    year = "2020"
}

@article{Suzuki:2021ium,
    author = "Suzuki, Hiromasa and Bamba, Aya and Shibata, Shinpei",
    title = "{Quantitative Age Estimation of Supernova Remnants and Associated Pulsars}",
    eprint = "2104.10052",
    archivePrefix = "arXiv",
    primaryClass = "astro-ph.HE",
    doi = "10.3847/1538-4357/abfb02",
    journal = "Astrophys. J.",
    volume = "914",
    number = "2",
    pages = "103",
    year = "2021"
}

@article{Zharikov:2021llh,
    author = "Zharikov, S. and Zyuzin, D. and Shibanov, Yu. and Kirichenko, A. and Mennickent, R. E. and Geier, S. and Cabrera-Lavers, A.",
    title = "{PSR B0656+14: the unified outlook from the infrared to X-rays}",
    eprint = "2101.07459",
    archivePrefix = "arXiv",
    primaryClass = "astro-ph.HE",
    doi = "10.1093/mnras/stab157",
    journal = "Mon. Not. Roy. Astron. Soc.",
    volume = "502",
    number = "2",
    pages = "2005--2022",
    year = "2021"
}

@article{Negreiros:2018cho,
    author = "Negreiros, Rodrigo and Tolos, Laura and Centelles, Mario and Ramos, Angels and Dexheimer, Veronica",
    title = "{Cooling of Small and Massive Hyperonic Stars}",
    eprint = "1804.00334",
    archivePrefix = "arXiv",
    primaryClass = "astro-ph.HE",
    doi = "10.3847/1538-4357/aad049",
    journal = "Astrophys. J.",
    volume = "863",
    number = "1",
    pages = "104",
    year = "2018"
}

@misc{PotekhinCoolData,
  author = {{Potekhin, A. Y.}},
  title = {Neutron Star Cooling Data},
  year = {2023},
  url = {https://www.ioffe.ru/astro/NSG/thermal/cooldat.html},
}

@article{Mauviard:2025dmd,
    author = "Mauviard, Lucien and others",
    title = "{A NICER View of the 1.4 M$_{⊙}$ Edge-on Pulsar PSR J0614-3329}",
    eprint = "2506.14883",
    archivePrefix = "arXiv",
    primaryClass = "astro-ph.HE",
    doi = "10.3847/1538-4357/ae145d",
    journal = "Astrophys. J.",
    volume = "995",
    number = "1",
    pages = "60",
    year = "2025"
}

@article{Salmi:2024bss,
    author = "Salmi, Tuomo and others",
    title = "{A NICER View of PSR J1231{\ensuremath{-}}1411: A Complex Case}",
    eprint = "2409.14923",
    archivePrefix = "arXiv",
    primaryClass = "astro-ph.HE",
    doi = "10.3847/1538-4357/ad81d2",
    journal = "Astrophys. J.",
    volume = "976",
    number = "1",
    pages = "58",
    year = "2024"
}

@article{Vinciguerra:2023qxq,
    author = "Vinciguerra, Serena and others",
    title = "{An Updated Mass{\textendash}Radius Analysis of the 2017{\textendash}2018 NICER Data Set of PSR J0030+0451}",
    eprint = "2308.09469",
    archivePrefix = "arXiv",
    primaryClass = "astro-ph.HE",
    doi = "10.3847/1538-4357/acfb83",
    journal = "Astrophys. J.",
    volume = "961",
    number = "1",
    pages = "62",
    year = "2024"
}

\bsp	
\label{lastpage}
\end{document}